\documentclass[10pt,final,journal,a4paper,oneside,twocolumn]{IEEEtran}
\usepackage{cite}

\usepackage{amsmath,amsfonts,amssymb}
\usepackage{xcolor}
\usepackage{mathdots}
\usepackage{bbm}
\usepackage{mathtools, cuted}
\usepackage{amssymb,bm}
\usepackage{tabularx}
\usepackage{booktabs}
\usepackage{multirow}
\usepackage{tikz}
\usetikzlibrary{calc,positioning}
\usepackage{overpic}
\usepackage{float}

\makeatletter
\let\MYcaption\@makecaption
\makeatother
\usepackage[font=footnotesize]{subcaption}
\makeatletter
\let\@makecaption\MYcaption
\makeatother

\usepackage{bbm}
\usepackage{mathtools, cuted}
\usepackage{amssymb,bm}
\usepackage{hyperref}
\usepackage{ragged2e}
\usepackage{overpic}
\usepackage{color}
\usepackage[super]{nth}
\usepackage{textcomp}

\usepackage{tikz}
\usetikzlibrary{calc,positioning}
\usepackage{array}
\newcolumntype{R}{>{$}r<{$}} 
\newcolumntype{C}{>{$}c<{$}} 
\usepackage[crop=pdfcrop]{pstool}
\usepackage{cancel}
\usepackage{dblfloatfix} 
\usepackage{tabularx}
\usepackage[usestackEOL]{stackengine}
\usepackage{blindtext}
\usepackage{siunitx}
\usepackage{tikz}
\usetikzlibrary{shapes.geometric, arrows,automata,positioning,shapes.multipart}
\usepackage{tabularx,colortbl}

\usepackage[super]{nth}

\usetikzlibrary{fit,backgrounds} 

\newcommand{\htext}[1]{%
	\makebox[0pt]{\Centerstack{#1}}
}

\newcommand{\vtext}[1]{%
	\makebox[0pt]{\rotatebox[origin=c]{90}{\Centerstack{#1}}}
}

\newcommand{\Et}{\tilde{E}}
\newcommand{\Ht}{\tilde{H}}
\newcommand{\Pt}{\tilde{P}}
\newcommand{\Mt}{\tilde{M}}

\newcommand{\Etf}{\tilde{\mathcal{E}}}
\newcommand{\Htf}{\tilde{\mathcal{H}}}
\newcommand{\Ptf}{\tilde{\mathcal{P}}}
\newcommand{\Mtf}{\tilde{\mathcal{M}}}

\newcommand{\E}{\mathbf{E}}

\newcommand{\J}{\mathbf{J}}
\newcommand{\K}{\mathbf{K}}

\renewcommand{\r}{\mathbf{r}}

\newcommand{\F}{\mathbf{F}}

\renewcommand{\S}{\mathbf{S}}

\newcommand{\A}{\mathbf{A}}
\renewcommand{\F}{\mathbf{F}}

\newcommand{\bbmatrix}{\begin{bmatrix}}
\newcommand{\ebmatrix}{\end{bmatrix}}

\newcommand{\chia}[2]{\chi_\text{#1}^{#2}}

\newcommand{\chib}[2]{\bar{\chi}_\text{#1}^{#2}}

\newcommand{\kv}{\mathbf{k}}

\newcommand{\Efd}{\tilde{\mathbf{{E}}}}

\newcommand{\Hfd}{\tilde{\mathbf{{H}}}}
\newcommand{\Pfd}{\tilde{\mathbf{{P}}}}
\newcommand{\Mfd}{\tilde{\mathbf{{M}}}}
\newcommand{\Jfd}{\tilde{\mathbf{{J}}}}
\newcommand{\Kfd}{\tilde{\mathbf{{K}}}}

\newcommand{\Esfd}{\tilde{\boldsymbol{\mathcal{E}}}}

\newcommand{\Hsfd}{\tilde{\boldsymbol{\mathcal{H}}}}
\newcommand{\Psfd}{\tilde{\boldsymbol{\mathcal{P}}}}
\newcommand{\Msfd}{\tilde{\boldsymbol{\mathcal{M}}}}
\newcommand{\Jsfd}{\tilde{\boldsymbol{\mathcal{J}}}}
\newcommand{\Ksfd}{\tilde{\boldsymbol{\mathcal{K}}}}

\newcommand{\Tsfd}{{{\mathcal{T}}}}
\newcommand{\Rsfd}{{{\mathcal{R}}}}

\newcommand{\Nh}{\mathbf{\hat n}}

\newcommand{\chit}{\overline{\overline{\chi}}}

\newcommand{\ee}{\text{ee}}

\newcommand{\zz}{{zz}}

\newcommand{\yy}{{yy}}

\renewcommand{\tt}{{yy}}

\newcommand{\mm}{\text{mm}}

\usepackage{empheq}

\definecolor{burntorange}{rgb}{0.8, 0.28, 0.0}
\definecolor{myGreen}{rgb}{0.0, 0.5, 0.0}
\definecolor{amber}{rgb}{0.8, 0.28, 0.0}
\definecolor{ceruleanblue}{rgb}{0.16, 0.28, 0.75}

\definecolor{ao}{rgb}{0.0, 0.5, 0.0}
\definecolor{cobalt}{rgb}{0.0, 0.28, 0.67}
\definecolor{amber}{rgb}{0.8, 0.36, 0.27}
\definecolor{ltblu}{rgb}{0.23, 0.27, 0.29}

\usepackage{balance}
\begin{document}
 \title{Part 1: Spatially Dispersive Metasurfaces: \\ Zero Thickness Surface Susceptibilities \\ \& Extended GSTCs}
\date{}
\author{Jo\~{a}o G. Nizer Rahmeier, \IEEEmembership{Student Member, IEEE}, Tom J. Smy, Jordan Dugan\\ and Shulabh Gupta \IEEEmembership{Senior Member, IEEE}
\thanks{ João G. Nizer Rahmeier, Tom J. Smy, Jordan Dugan, and Shulabh Gupta are with Carleton University, Ottawa, Canada (e-mail: JoaoNizer@cmail.carleton.ca). }}

\maketitle

\begin{abstract}
A simple method to describe spatially dispersive metasurfaces is proposed where the angle-dependent surface susceptibilities are explicitly used to formulate the zero thickness sheet model of practical metasurface structures. It is shown that if the surface susceptibilities of a given metasurface are expressed as a ratio of two polynomials of tangential spatial frequencies, $\boldsymbol{k_{||}}$ with complex coefficients, they can be conveniently expressed as spatial derivatives of the difference and average fields around the metasurface in the space domain, leading to extended forms of the standard Generalized Sheet Transition Conditions (GSTCs) accounting for the spatial dispersion. Using two simple examples of a short electric dipole and an all-dielectric cylindrical puck unit cells, which exhibit purely tangential surface susceptibilities and reciprocal/symmetric transmission and reflection characteristics, the proposed concept is numerically confirmed in 2D. A single Lorentzian has been found to describe the spatio-temporal frequency behavior of a short dipole unit cell, while a multi-Lorentzian description is developed to capture the complex multiple angular resonances of the dielectric puck. For both cases, the appropriate spatial boundary conditions are derived. 
\end{abstract}



\section{Introduction}

Electromagnetic Metasurfaces are 2D arrays of sub-wavelength resonating particles that derive their macroscopic field response from their geometry and material arrangements at the microscopic scale~\cite{Metasurface_Review}. By engineering these resonating particles, a wide variety of macroscopic fields transformations may be achieved, which has led to a myriad of exotic applications across the electromagnetic spectrum, ranging from cloaking, illusions to holograms extended to real-time reconfiguration of wireless environments~\cite{Optical_MS_Reconfig, Reconfigr_MS, CloakingReview, MS_review_Yu, Smy_Close_ILL, smy2020IllOpen, Fink_AI_Metasurface}.

From their very nature, metasurfaces have a multi-scale architecture with sub-wavelength resonators arranged in an electrically large array. Moreover, computation of the scattered fields from these metasurfaces inherently needs to resolve the sub-wavelength features, which is generally a significant computational task -- affecting both design synthesis and subsequent field analysis of electrically large metasurfaces. As a result, exploiting their electrically thin characteristics, they have been modeled as zero thickness surfaces, as spatial discontinuities, described in terms of dipolar tensorial electric and magnetic surface susceptibilities, $\bar{\bar{\chi}}(\omega, \r)$ \cite{GenBCEM, MS_Synthesis}. The Generalized Sheet Transition Conditions govern the resulting macroscopic fields (GSTCs)~\cite{KuesterGSTC,IdemenDiscont} involving electric and magnetic surface polarization densities.

An essential step in building equivalent zero thickness models of practical metasurfaces is mapping geometrical/material characteristics to tensorial surface susceptibilities at specified design frequencies, accounting for a complete angular scattering of the surface beyond paraxial wave propagation. Most of the work in the literature has been focussed on one class of these structures: \emph{spatially non-dispersive metasurfaces}, where the surface polarizations are induced due to local field interaction only. In such cases, the surface susceptibilities, being characteristic properties of the metasurface, are entirely independent of the angle of incidence of the incoming waves. As a result, the angular scattering behavior of the metasurfaces can thus be described in terms of the normal components of the surface polarizations, assuming they are physically supported by the structure~\cite{Karim_Angular_MS,Karim_Bianiso_MS, GenBCEM}.

Owing to a virtually unlimited number of metasurface structures proposed and demonstrated in the general area of electromagnetic metasurfaces, the foundational resonators structures range from deeply sub-wavelength sizes to close to wavelength dimensions. Of which a typical example, out of many, are all-dielectric resonator-based Huygens' structures, for instance~\cite{Cylindrical_DMS, West_DMS_Lens, AllDieelctricMTMS, Soichi_Huygen_mmWave}. Thus, a priori, metasurface structures are not necessarily spatially non-dispersive and may exhibit a \emph{non-local response}, i.e., spatial dispersion~\cite{Asadchy2017}. Such structures, in general, cannot be modeled using standard dipolar surface susceptibility models and, in particular, the normal surface polarization alone.

Some recent works have explored surface susceptibility models of spatially dispersive metasurfaces, particularly in terms of multipolar description of the resonators and hyper-surface susceptibilities~\cite{Albooyeh_SD, achouri2021multipolar, achouri2021extension}. However, these techniques have been typically demonstrated for weak spatial dispersion and have not replicated the complete angular scattering response of an arbitrary metasurface, which may even exhibit multiple resonances across the angular spectrum. Therefore, a general zero thickness modeling of spatially dispersive structures, particularly for strong dispersion, has been an open problem in the literature. In the present work, we seek to describe general metasurface structures in terms of a compact surface susceptibility model compatible with the GSTCs, which represents general spatial boundary conditions.

This paper represents Part~1 of this work, where we propose a simple angle-dependent surface susceptibility modeling technique that forms an equivalent zero thickness sheet model of an arbitrary spatially dispersive metasurface structure. We specifically exploit angle-dependent surface susceptibilities and describe them as a ratio of polynomials of the transverse wave-vector $k_{||}$. We further describe them in terms of physically motivated Lorentz oscillator model with angle-dependent resonator parameters, which leads to a very compact way to describe the complete angular scattering of the metasurfaces. Moreover, they feature convenient spatial domain forms with spatial derivatives of both the difference and average fields at the metasurface, resulting in extended GSTCs, which can be straightforwardly integrated into electromagnetic fields solvers as generalized boundary conditions, e.g., an Integral Equation (IE) solver, as presented in Part~II of this work~\cite{Part_2_Smy_SD}. 

The paper is structured as follows. Sec.~II reviews the conventional GSTCs and proposes the principle of modeling an arbitrary spatially dispersive metasurface structure based on angle-dependent surface susceptibilities. A general representation of angle-dependent surface susceptibilities as a ratio of two polynomials is proposed in Sec.~III, which leads to the extended form of the GSTCs involving spatial derivatives of difference and average fields at the metasurface. Next, a Lorentz oscillator model is proposed, which forms a compact model to represent arbitrarily complex structures with multiple angular resonances. The proposed technique is then demonstrated numerically in Sec.~IV for two structures: short dipoles and an all-dielectric resonator. Sec~V further discusses a connection between spatial dispersion and the normal surface susceptibility components. Finally, conclusions are provided in Sec.~VI, summarizing the work and describing the future steps.

\section{Spatially Dispersive Metasurfaces}

\subsection{Angle Dependent Surface Susceptibilities}

\begin{figure}[tbp]
\begin{center}
   	 \begin{overpic}[width=0.9\linewidth,grid=false,trim={0cm 0cm 0cm 0cm},clip]{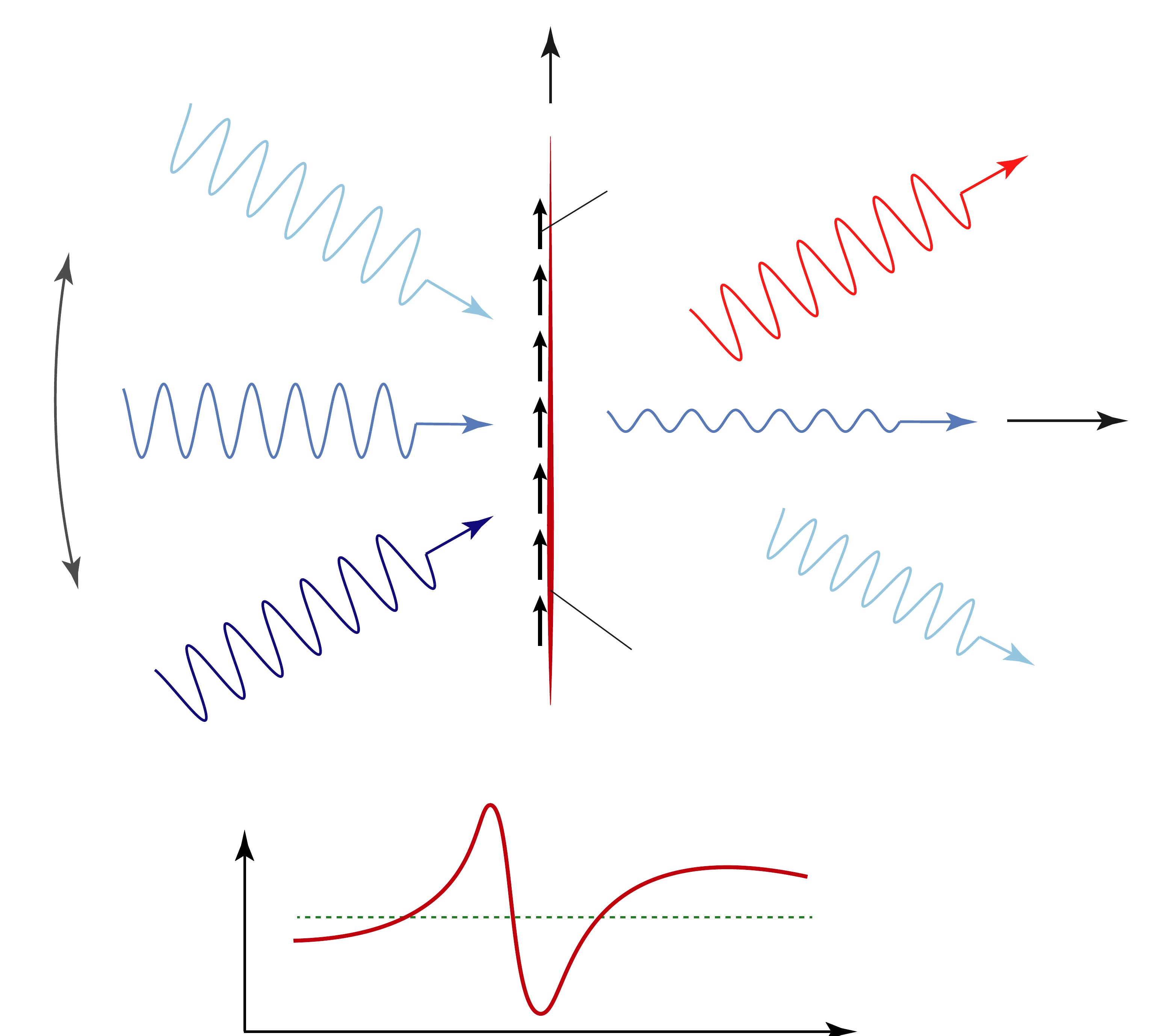}
            \put(98,  52){\scriptsize \htext{ $x$}}
            \put(47,  87){\scriptsize \htext{ $y$}}
            \put(7,  26){\scriptsize\htext{ $\psi_0(\theta_1,\omega)$}}\put(7,  75){\scriptsize\htext{ $\psi_0(\theta_3,\omega)$}} \put(25,  57){\scriptsize\htext{ $\psi_0(\theta_2,\omega)$}}
            \put(85,  78){\scriptsize\htext{ $\psi_t(\theta_3,\omega)$}}\put(93,  28){\scriptsize\htext{ $\psi_t(\theta_1,\omega)$}}\put(77,  48){\scriptsize\htext{ $\psi_t(\theta_2,\omega)$}}
            \put(2,  52){\scriptsize\vtext{ Angle of Incidence, $\theta$}} 
            \put(79,  84){\htext{\shortstack{\color{amber}\footnotesize \textbf{\textsc{Dispersive Metasurface}} \\ \scriptsize Angle Dependent Susceptibilities, $\bar{\bar{\chi}}(\theta)$}}}
             \put(20,  84){\scriptsize\htext{\color{ltblu}\shortstack{ Incident Plane Waves \\ e.g. TE Mode ($E_z,~H_x,~H_z$)}} }
              \put(47,  26){\scriptsize \htext{ $x = 0$}} \put(65, 27){\scriptsize \htext{ Uniform Surface}}
              \put(60,  66){\scriptsize \color{ao} \htext{ $\bar{\bar{\chi}}(\theta_3)$}}
              \put(54,  48){\scriptsize \color{ao}\htext{ $\bar{\bar{\chi}}(\theta_2)$}}
              \put(60,  40){\scriptsize \color{ao}\htext{ $\bar{\bar{\chi}}(\theta_1)$}}
              \put(60,  72){\scriptsize\htext{ $\{\J_s,~\K_s\}$}}
               \put(85,  -1){\scriptsize\htext{ $k_y = k_0\sin\theta$}}  \put(45,  -4){\scriptsize\htext{angular spectrum }}
               \put(22,  19){\scriptsize\htext{ $\chi_{\alpha,\beta}$}}
               \put(82,  6){\scriptsize\htext{ \color{ao}\shortstack{Non-dispersive \\ $\chi_{\alpha,\beta}(k_y) = \text{const}$}}}
    \end{overpic}
\end{center}
\caption{The general field problem of a spatially dispersive metasurface where its surface susceptibilities, $\bar{\bar{\chi}}(\theta)$ are dependent on the angle of incidence, $\theta$ of the incoming plane-waves. A uniform surface with purely tangential surface susceptibilities with no bi-anisotropic terms, excited with TE mode incidence fields are assumed throughout this work for simplicity.}\label{fig:Problem}
\end{figure}
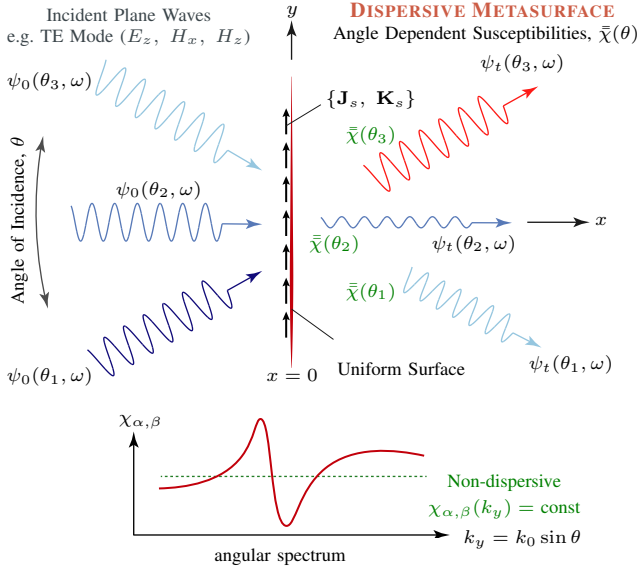

Consider a field scattering problem from a metasurface which is lying in the $y$-$z$ plane, and incident with arbitrary fields, thereby generating the scattered fields in the transmission and reflection regions, as shown in Fig.~\ref{fig:Problem}. The metasurface is assumed to be a zero thickness sheet, i.e., $\delta = 0$. When excited with an incident plane-wave, $\psi_0(\theta, \omega)$, induced electric and magnetic surface currents, $\{\J_s,~\K_s\}$ are generated, which then re-radiate to produce the scattered fields in both reflection and transmission regions. For spatially non-dispersive metasurfaces, these currents represent the surface response to the local incident fields only (i.e., point-by-point interaction). Thus, their zero thickness sheet model can be described using constant (independent of the angle of incidence, $\theta$) electric and magnetic dipolar surface susceptibility tensors, $\bar{\bar{\chi}}$. Hence, the corresponding dipolar surface polarizations are related to the averaged fields around the metasurface~\cite{Xiao:2019aa}, as
\begin{subequations}\label{Eq:PM_NSD}
\begin{align}
    \Pfd &= \epsilon_0 \chit_\ee \Efd_\text{av} + \frac{1}{c_0} \chit_\text{em} \Efd_\text{av}\\
    \Mfd &= \mu_0 \chit_\mm \Hfd_\text{av} + \frac{1}{\eta_0} \chit_\text{me} \Efd_\text{av}
\end{align}
\end{subequations}
where each of the tensors $\bar{\bar{\chi}}_{\alpha, \beta}$ are $(3\times 3)$ matrices containing both tangential and normal susceptibility components. In addition, the electromagnetic fields at the metasurface follow the Generalized Sheet Transition Conditions (GSTCs) given by\footnote{The various fields and field components are represented as $\E(\r,t)$ for the space and time domains, $\Efd(\r, \omega)$ for the temporal frequency domain, $\boldsymbol{\mathcal{E}}(\kv,t)$ for the spatial frequency domain, and $\Esfd(\kv,\omega)$ for the spatial and temporal frequency domains. Time convention used is $e^{j\omega t}$.},
\begin{subequations}\label{Eq:ConvGSTC}
\begin{align}
	\Delta \Efd &= j\omega (\Nh \times  \Mfd) - \nabla_{T}\left(\frac{\Pfd_{n}}{\epsilon}\right)\\
	\Delta \Hfd  &= -j\omega (\Nh \times\Pfd) - \nabla_{T}\left(\frac{\Mfd_{n}}{\mu}\right)
\end{align}
\end{subequations}
where $\Nh$ is the surface normal (e.g. along $x-$axis) and $\Delta\psi = (\psi^+ - \psi^-)$ is the field difference across the metasurface \cite{KuesterGSTC, Metasurface_Review, IdemenDiscont}. 

However, for a spatially dispersive metasurface, the surface susceptibilities are functions of the incoming incidence angle (or spatial frequencies, $k_y$), as illustrated in Fig.~\ref{fig:Problem}. This is analogous to temporal dispersion where the constitutive parameters of a medium depend on the temporal frequencies, i.e. $\epsilon = \epsilon(\omega)$ and $\mu = \mu(\omega)$. It results in a re-arrangement of the instantaneous temporal frequencies when a broadband signal propagates through the medium, leading to envelope distortion in time. For spatially dispersive surfaces, the induced surface currents depend on $\theta$, so that their constitutive parameters, $\bar{\bar{\chi}}_{\alpha, \beta} = \bar{\bar{\chi}}_{\alpha, \beta}(k_y)$. This means that the induced currents at a location $y$ on the surface depend on the local, as well as non-local fields across the metasurface so that a point-by-point interaction no longer holds. Moreover, the spatial frequencies of an incoming \emph{spatially broadband} signal will be re-arranged in space, distorting the spatial field envelope. Therefore, an engineered spatially dispersive surface acts like a \emph{spatial frequency filter}.

For simplicity, let us consider a symmetric structure with TE mode excitation, so that  $\bar{\bar{\chi}}_\text{em, me}$ terms are zero, and we further assume that there are no normal components, so that the surface is completely described by tangential surface susceptibilities, i.e., $\tilde{\chi}_\text{mm}^{yy}$ and $\tilde{\chi}_\text{ee}^{zz}$ \cite{Karim_Angular_MS, Karim_Bianiso_MS}. Using \eqref{Eq:PM_NSD} and \eqref{Eq:ConvGSTC}, we can compute the two surface susceptibilities as \cite{Metasurface_Synthesis_Caloz}
\begin{subequations}\label{Eq:TangChi}
\begin{align}
\tilde{\chi}_\text{ee}(k_y=k_0\sin\theta) &= \frac{2j\cos\theta}{k_0}\left(\frac{\Rsfd+\Tsfd-1}{\Tsfd+\Rsfd+1}\right)\\
\tilde{\chi}_\text{mm}(k_y=k_0\sin\theta) &= \frac{2j}{k_0\cos\theta}\left(\frac{\Rsfd-\Tsfd+1}{\Rsfd-\Tsfd -1}\right)
\end{align}
\end{subequations}
where $\Rsfd$ and $\Tsfd$ are the reflection and transmission response of the surface, as a function of the angle of incidence $\theta$ of an incoming uniform plane-wave, i.e. $\delta(k_y - k_0\sin\theta)$, at a specified temporal frequency $\omega$. Under these conditions of non-local interaction, \eqref{Eq:PM_NSD} does not hold, and a relationship between induced surface currents and the average fields, via the angle-dependent surface susceptibilities, must be modified to account for the non-local field interaction.

\subsection{Surface Polarizations}
 
In order to avoid mathematical complexity, and focus on the underlying physical concept, consider a uniform metasurface, lying in the $y-z$ plane at $x=0$, and excited with an oblique incident uniform plane-wave, which induces surface currents $\Jfd$ and $\Kfd$ on the surface. The fields radiated by these currents are obtained using the electric and magnetic vector potentials $\A(\r)$ and $\F(\r)$~\cite{Rothwell:2018aa}. Since the currents are only on the surface and are fully tangential, $\Jfd(y) = \Jfd_\text{s}\delta(y')$, so that \cite{Albooyeh_SD}
\begin{align}
\A(\r) =  -\frac{j\mu}{2k} \Jfd_\text{s} e^{-jk|x|},~\F(\r) = -\frac{j\epsilon}{2k} \Kfd_\text{s}e^{-jk|x|}.
\end{align}
\noindent The scattered fields radiated due to these currents may be obtained as
\begin{subequations}\label{eq:GenScat}
\begin{align}
\Efd_s &= - \frac{\eta}{2} \Jfd_\text{s} e^{-jk|x|}  - \frac{j}{2k}\nabla\times \Kfd_\text{s}e^{-jk|x|} \\
\Hfd_s &= - \frac{1}{2\eta} \Kfd_\text{s} e^{-jk|x|}  - \frac{j}{2k}\nabla\times \Jfd_\text{s} e^{-jk|x|} 
\end{align}
\end{subequations}
\noindent  The total fields on each side of the surface are then given by the sum of the incident and total scattered fields due to these currents. Assuming incident fields only on the reflection side of the metasurface ($x<0$), the average fields are given by
\begin{subequations}\label{Eq:AvK}
\begin{align}
 \Efd_\text{av} &= \frac{\Efd^+_\text{s} + (\Efd^-_\text{s} + \Efd_\text{i})}{2} =   - \frac{\eta_0}{2} \Jfd_\text{s} +\frac{\Efd_\text{i}}{2}\\
 \Hfd_\text{av} &= \frac{\Hfd^+_\text{s} + (\Hfd^-_\text{s} + \Hfd_\text{i})}{2} =    - \frac{1}{2\eta_0} \Kfd_\text{s} + \frac{\Hfd_\text{i}}{2}
\end{align}
\end{subequations}
For \emph{local field interaction}, the incident fields are related to the surface currents at $x=0$, as:
\begin{subequations}\label{Eq:local}
\begin{align}
\Jfd_\text{s}(y) &= \bar{\bar{\alpha}}_\text{ee}'(y)\cdot \Efd_\text{i}(y) +  \bar{\bar{\alpha}}_\text{em}'(y)\cdot \Hfd_\text{i}(y)\\
\Kfd_\text{s}(y) &= \bar{\bar{\alpha}}_\text{mm}'(y)\cdot \Hfd_\text{i}(y) +  \bar{\bar{\alpha}}_\text{me}'(y)\cdot \Efd_\text{i}(y)
\end{align}
\end{subequations}
where it is assumed that $\bar{\bar{\alpha}}'_{\alpha, \beta}$ are angle-independent, and thus non-dispersive. The currents are thus induced due to local fields only following point-by-point interaction. However, for a general \emph{spatially-dispersive metasurface, with non-local field interaction}, the surface fields are given by the spatial convolution of the average fields with the polarization response of the surface, i.e.
\begin{subequations}
\begin{align}
\Jfd_\text{s}(y) &= \bar{\bar{\alpha}}_\text{ee}'(y)\ast \Efd_\text{i}(y) +  \bar{\bar{\alpha}}_\text{em}'(y)\ast \Hfd_\text{i}(y)\\
\Kfd_\text{s}(y) &= \bar{\bar{\alpha}}_\text{mm}'(y)\ast \Hfd_\text{i}(y) +  \bar{\bar{\alpha}}_\text{me}'(y)\ast \Efd_\text{i}(y)
\end{align}
\end{subequations}
To avoid convolution operation in space, this can easily be expressed in the spatial frequency domain, i.e., $\mathcal{F}_y\{\psi(y)\} =\psi(k_y)$, so that,
\begin{subequations}
\begin{align}
\Jsfd_\text{s}(k_y) &= \bar{\bar{\alpha}}_\text{ee}'(k_y)\cdot \Esfd_\text{i}(k_y) +  \bar{\bar{\alpha}}_\text{em}'(k_y)\cdot \Hsfd_\text{i}(k_y)\\
\Ksfd_\text{s}(k_y) &= \bar{\bar{\alpha}}_\text{mm}'(k_y)\cdot \Hsfd_\text{i}(k_y) +  \bar{\bar{\alpha}}_\text{me}'(k_y)\cdot \Esfd_\text{i}(k_y)
\end{align}
\end{subequations}
where $k_y$ represents the tangential component of the wave-vector $\kv$ along the surface (i.e., related to the incidence angle of the incoming wave). Isolating $\Esfd_\text{i}$ and $\Hsfd_\text{i}$, substituting them in the spatial Fourier transform of \eqref{Eq:AvK}, and rearranging the terms, we relate the average fields to the surface currents, i.e.
\begin{subequations}
\begin{align}
\Esfd_\text{av}(k_y) &= \bar{\bar{\zeta}}_\text{ee}(k_y)\cdot \Jsfd_\text{s}(k_y)  +  \bar{\bar{\zeta}}_\text{em}(k_y)\cdot \Ksfd_\text{s}(k_y)\\
\Hsfd_\text{av}(k_y)  &= \bar{\bar{\zeta}}_\text{mm}(k_y)\cdot \Jsfd_\text{s}(k_y)  +  \bar{\bar{\zeta}}_\text{me}(k_y)\cdot \Ksfd_\text{s}(k_y)
\end{align}
\end{subequations}
Finally, expressing surface dipole moments, as surface current densities over the unit cell area $\S$, as $\Jsfd_\text{s} = j\omega\Psfd$ and $\Ksfd_\text{s} =  j\omega\Msfd$, and solving for the surface polarizations, we get
\begin{subequations}\label{Eq:Pky}
\begin{align}
\Psfd(k_y) = \epsilon_0\bar{\bar{\chi}}_\text{ee}(k_y)\cdot  \Esfd_\text{av}(k_y)  +  \frac{1}{c_0}\bar{\bar{\chi}}_\text{em}(k_y)\cdot  \Hsfd_\text{av}(k_y)\\
\Msfd(k_y)  = \bar{\bar{\chi}}_\text{mm}(k_y)\cdot  \Hsfd_\text{av}(k_y)  +  \frac{1}{\eta_0}\bar{\bar{\chi}}_\text{me}(k_y)\cdot  \Esfd_\text{av}(k_y)
\end{align}
\end{subequations}
where $\bar{\bar{\chi}}_{\alpha\beta}$ are the \emph{angle-dependent} surface susceptibilities. For a non-spatially dispersive metasurface with local field interaction only, $\bar{\bar{\chi}}(k_y) = \text{const.}$, so that we retrieve the standard surface susceptibility relations of \eqref{Eq:PM_NSD} following \eqref{Eq:local}.

 \section{Extended Generalized Sheet Transition Conditions (GSTCs)}

\subsection{Susceptibilities as Ratio of Partial Fractions}

Let us consider a spatially dispersive metasurface that is characterized using tangential surface susceptibilities only, i.e. $\chi_\text{ee}^{zz}$ and $\chi_\text{mm}^{yy}$. Using \eqref{Eq:Pky}, for a TE mode, we get
\begin{subequations}\label{Eq:PMky}
\begin{align}
\Ptf_z(k_y) &= \epsilon_0\tilde{\chi}_\ee^{\zz}(k_y)\cdot \Etf_{z,\text{av}}(k_y)\\
\Mtf_y(k_y) &=\tilde{\chi}_\text{mm}^{yy}(k_y)\cdot\Htf_{y,\text{av}}(k_y)
\end{align}
\end{subequations}
To model such a metasurface, we need to express their angular dependence using a convenient functional representation. We postulate that a general surface susceptibility function can be expressed as a ratio of two polynomials in $k_y$, with known complex coefficients featuring various possible poles and zeros \cite{Nizer_ComplexModes} accounting for angular resonances, so that we can write
\begin{subequations}\label{Eq:Ratio}
\begin{align}
\tilde{\chi}_\ee^{\zz}(k_y) &= \left(\frac{\sum_m a_m k_y^m}{\sum_n b_n k_y^n}\right) \\
\tilde{\chi}_\text{mm}^{yy}(k_y) &= \left(\frac{\sum_m c_m k_y^m}{\sum_n d_n k_y^n}\right).
\end{align}
\end{subequations}
Using \eqref{Eq:PMky}, this can be expressed as:
\begin{subequations}
\begin{align}
\sum_n b_n k_y^n \Ptf_z(k_y) = \epsilon_0\sum_m a_m k_y^m \Etf_{z,\text{av}}(k_y)\\
\sum_n d_n k_y^n \Mtf_y(k_y) = \sum_m c_m k_y^m \Htf_{x,\text{av}}(k_y)
\end{align}
\end{subequations}
 Taking the \emph{inverse spatial Fourier transform}, the terms with various polynomial orders turn into spatial derivatives, resulting in
\begin{subequations}
\begin{align}
\sum_n b_n j^n  \frac{\partial^n \tilde{P}_z}{\partial y^n} = \epsilon_0\sum_m a_m j^m  \frac{\partial^m \tilde{E}_{z,\text{av}}}{\partial y^m}\\
\sum_n d_n j^n  \frac{\partial^n \tilde{M}_y}{\partial y^n} = \sum_m c_m j^m  \frac{\partial^m \tilde{H}_{x,\text{av}}}{\partial y^m}  
\end{align}
\end{subequations}
Now, we know from the GSTCs that a portion of the field differences can be associated with this polarization, i.e.
\begin{subequations}\label{Eq:GSTCp1}
\begin{align}
\Delta \tilde{H}_x &= j\omega\tilde{P}_z\\
\Delta \tilde{E}_z &= j\omega \tilde{M}_y
\end{align}
\end{subequations}
 Substituting this in the above equation results in an extended form of the GSTC to
\begin{subequations}\label{Eq:ExtendedGSTCs}
\begin{align}
\sum_n b_n j^n  \frac{\partial^n \Delta \tilde{H}_y}{\partial y^n} &= j\omega\epsilon_0 \sum_m a_m j^m  \frac{\partial^m \tilde{E}_{z,\text{av}}}{\partial y^m}\\
\sum_n d_n j^n  \frac{\partial^n \Delta \tilde{E}_z}{\partial y^n} &= j\omega \sum_m c_m j^m  \frac{\partial^m \tilde{H}_{y,\text{av}}}{\partial y^m}  
\end{align}
\end{subequations}
Therefore, for a metasurface that is explicitly described using tangential surface susceptibilities only, spatial dispersion manifests as spatial derivatives of both the difference fields \emph{and} the average fields across the metasurface. This is the key result of this work.  For a spatially non-dispersive cell, the above extended GSTC naturally reduces to their standard forms as
\begin{subequations}
\begin{align}
\Delta \tilde{H}_y  &= j\omega  \epsilon_0 \chi_\ee^\zz  \tilde{E}_{z,\text{av}} \\
\Delta \tilde{E}_z  &= j\omega   \chi_\mm^{yy}  \tilde{H}_{y,\text{av}} 
\end{align}
\end{subequations}
with $b_0 = d_0 = 1$, $a_0 = \chi_\ee^\zz$ and $c_0= \chi_\mm^{yy}$. 
It should be noted that for a perfectly symmetric metasurface invariant to the sign of $\pm k_y$, only even orders of the polynomial exist in the general representation of \eqref{Eq:Ratio}, i.e., only even derivatives in \eqref{Eq:ExtendedGSTCs}. The extended GSTCs of \eqref{Eq:ExtendedGSTCs}, now represent general boundary conditions which can now be operated on arbitrary incident fields, and be integrated in various standard field solvers such as FDTD \cite{Caloz_MS_Siijm, Smy_Metasurface_Space_Time, Smy_FDTD, CalozFDTD} and Integral Equation (IE) methods \cite{Smy_Close_ILL, smy2020IllOpen}, for instance. The IE-GSTC implementation of \eqref{Eq:ExtendedGSTCs} is presented in Part 2 of this work~\cite{Part_2_Smy_SD}.

\subsection{Lorentz Oscillator Model}

To understand the angle dependent unit cell resonances captured by the poles of \eqref{Eq:Ratio}, and their physical origins, consider an arbitrary sub-wavelength unit cell structure excited by an oblique uniform plane wave (i.e. incidence angle $\theta$ or $k_y = k_0\sin\theta$), which can be described using a standard Lorentz oscillator model:
\begin{align}\label{Eq:TemporalLorentz}
\frac{\partial^2 \mathcal{P}_z}{\partial t^2} + \gamma(k_y) \frac{\partial \mathcal{P}_z}{\partial t}  + \omega_0^2(k_y)\mathcal{P}_z = \epsilon_0\omega_p^2(k_y)\mathcal{E}_{z,\text{av}}
\end{align}
which describes the temporal electric (and magnetic) surface polarization function in response to the average electric (and magnetic) fields around the surface, and where $\omega_p^2$ is the plasma frequency, $\gamma$ is a damping coefficient and $\omega_0$ is the resonant frequency at that specific angle of incidence. In the temporal frequency domain, we can express this using a temporal Fourier transform, $\mathcal{F}_t\{\cdot\}$ as: 
\begin{align}\label{Eq:Lor_kx_w}
-\omega^2\Ptf_z + j\gamma\omega \Ptf_z  + \omega_0^2\Ptf_z= \epsilon_0\omega_p^2\Etf_{z,\text{av}}
\end{align}
To account for the angular dependence of the unit cell, let us heuristically assume that the damping coefficient and the resonant frequency are polynomial functions of the incoming plane-wave angles such that 
\begin{subequations}\label{Eq:gamma_omega_kx}
\begin{align}
\gamma &= \alpha_0 + \alpha_1 k_y + \alpha_2 k_y^2 + O(k_y^n) \\
\omega_p^2 &= \beta_0^2 +\beta_1 k_y  + \beta_2 k_y^2 + O(k_y^n) \\
\omega_0^2 &= \zeta_0^2 +\zeta_1 k_y  + \zeta_2 k_y^2 + O(k_y^n)
\end{align}
\end{subequations}
Consider terms up to the second order for $\gamma$, $\omega_p^2$ and $\omega_0^2$. Then, substituting \eqref{Eq:gamma_omega_kx} in \eqref{Eq:Lor_kx_w}, allows us to relate the average fields with the polarization in the following form,
\begin{align}\label{Eq:Lor_kx_w_2}
\Ptf_z^z  = \epsilon_0 \frac{\chi^\zz_{\ee,0}+j\chi^\zz_{\ee,1}k_y-\chi^\zz_{\ee,2}k_y^2}{(1 + j\xi_{\ee,1}^\zz k_y - \xi_{\ee,2}^\zz k_y^2)}\Etf_{z,\text{av}}
\end{align}
 \noindent where,
 \begin{align*}
     \xi_{\ee,1}^\zz &= \left[\frac{(\omega \alpha_1 - j\zeta_1)}{(\zeta_0^2- \omega^2 + j\omega \alpha_0)}\right]\\
    \xi_{\ee,2}^\zz &= \left[\frac{-(\zeta_2 + j\omega \alpha_2)}{(\zeta_0^2- \omega^2 + j\omega \alpha_0)}\right]\\
    \chi^\zz_{\ee,0} &= \left[\frac{\beta_0^2}{(\zeta_0^2- \omega^2 + j\omega \alpha_0)}\right]\\
    \chi^\zz_{\ee,1} &= \left[\frac{-j\beta_1}{(\zeta_0^2- \omega^2 + j\omega \alpha_0)}\right]\\
    \chi^\zz_{\ee,2} &= \left[\frac{-\beta_2}{(\zeta_0^2- \omega^2 + j\omega \alpha_0)}\right].
\end{align*}
 %
 For the case of spatially symmetric unit cells that is assumed throughout this work, the terms $\xi_{1}$ in \eqref{Eq:Lor_kx_w_2} will be zero.
 Finally, taking an inverse spatial Fourier transform $\mathcal{F}_y^{-1}\{\cdot\}$ of \eqref{Eq:Lor_kx_w_2}, we get (along with the analogous equation for the magnetic surface polarization)
 \begin{subequations}\label{Eq:SpatialLorentz}
 \begin{align}
      \left[\xi_{\ee,2}^\zz \frac{\partial^2}{\partial y^2}  + 1\right]\Pt_z &= \epsilon_0 \left[\chi_{\ee,2}^\zz \frac{\partial^2}{\partial y^2} + \chi_{\ee,0}^\zz \right]\Et_{z,\text{av}},\\ 
      \left[\xi_{\mm,2}^\zz \frac{\partial^2}{\partial y^2} +  1\right]\Mt_z &=  \left[\chi_{\mm,2}^{yy} \frac{\partial^2}{\partial y^2} + \chi_{\mm,0}^{yy} \right]\Ht_{y,\text{av}}.  
 \end{align}
 \end{subequations}
which appears as a spatial counterpart of \eqref{Eq:TemporalLorentz} and represents the spatial boundary condition across the zero thickness sheet. While the general form of \eqref{Eq:Ratio} is applicable in general, the Lorentzian form of \eqref{Eq:Lor_kx_w_2} represents an important special case motivated by physical considerations.
 
\section{Application to Practical Metasurface Structures} 

To illustrate the proposed method, we will consider two example metasurfaces composed of a 2D array of a) a short electric dipole, and b) a cylindrical dielectric puck, respectively, lying in the $y$-$z$ plane, with $x$-$y$ as the plane-of-incidence and TE mode excitation. Both the unit cell structures exhibit symmetry about the origin so that their transmittance/reflectance is an even function of $\theta$ and have identical and reciprocal responses for left and right excitations. It can be shown that under these conditions, only $\chia{ee}{zz}$, $\chia{mm}{yy}$ and $\chia{mm}{xx}$ are the possible non-zero susceptibility tensor components. Furthermore, it can be shown that for both these structures, $\chia{mm}{xx}$ is also zero, so that these structures are completely described in terms of tangential surface susceptibilities and are thus birefringent~\cite{Karim_Angular_MS}. 

\subsection{Short Metal Dipole}

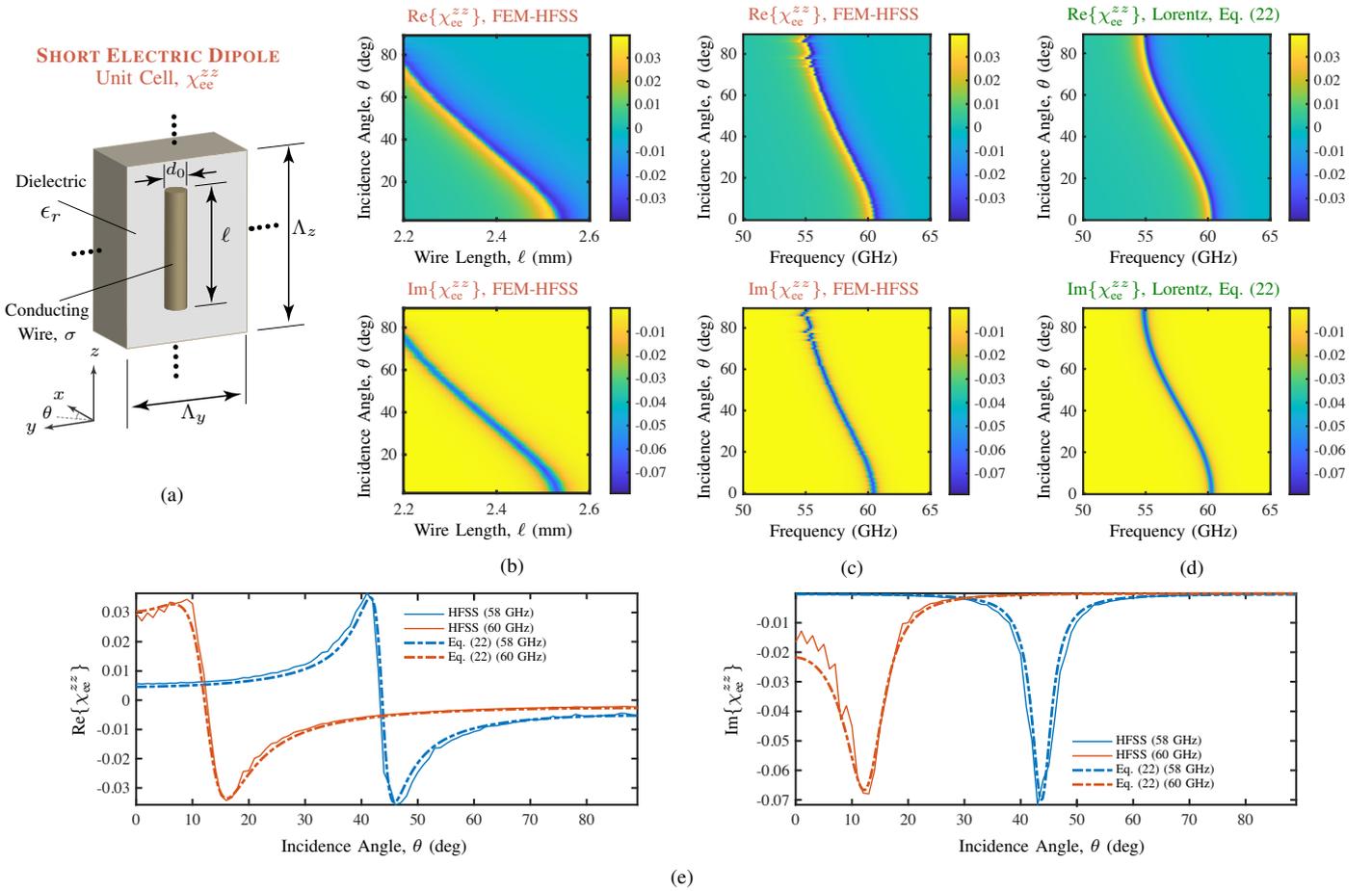
\begin{figure*}[!htbp]
		\centering
		\begin{subfigure}{0.475\columnwidth}
		\centering
					\begin{overpic}[width=\columnwidth,grid=false,trim={0cm 0cm 0cm 0cm},clip]{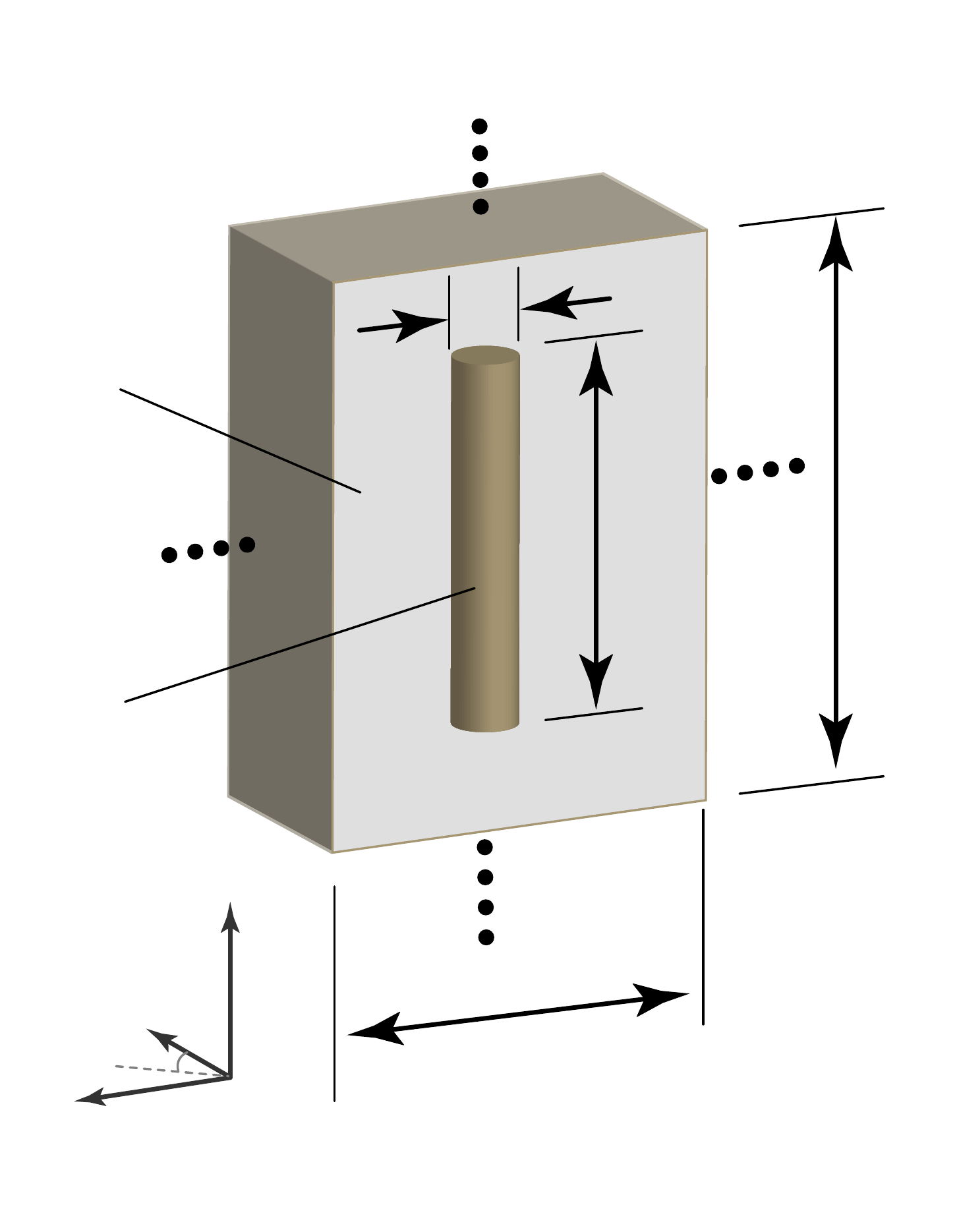}
				\put(52, 58){\htext{\footnotesize $\ell$}}
				\put(8, 68){\htext{\scriptsize Dielectric \\ $\epsilon_r$}}
				\put(7, 35){\htext{\scriptsize \shortstack{Conducting \\ Wire, $\sigma$}}}
				\put(3,10){\htext{\scriptsize $y$}}
				\put(10, 18){\htext{\scriptsize $x$}}
				\put(19, 29){\htext{\scriptsize $z$}}
				\put(7, 13){\htext{\scriptsize $\theta$}}
				\put(39.5, 75){\htext{\scriptsize $d_0$}}
				\put(72, 60){\htext{\footnotesize $\Lambda_z$}}
				\put(44, 13){\htext{\footnotesize $\Lambda_y$}}
				\put(35, 100){\htext{\footnotesize \color{amber}\shortstack{\textbf{\textsc{Short Electric Dipole}}\\ Unit Cell, $\chia{ee}{zz}$}}}
		\end{overpic} \caption{}
\end{subfigure}
\hfill
\begin{subfigure}{0.475\columnwidth}
				\begin{overpic}[width=\columnwidth,grid=false,trim={0cm 0cm 0cm 0cm},clip]{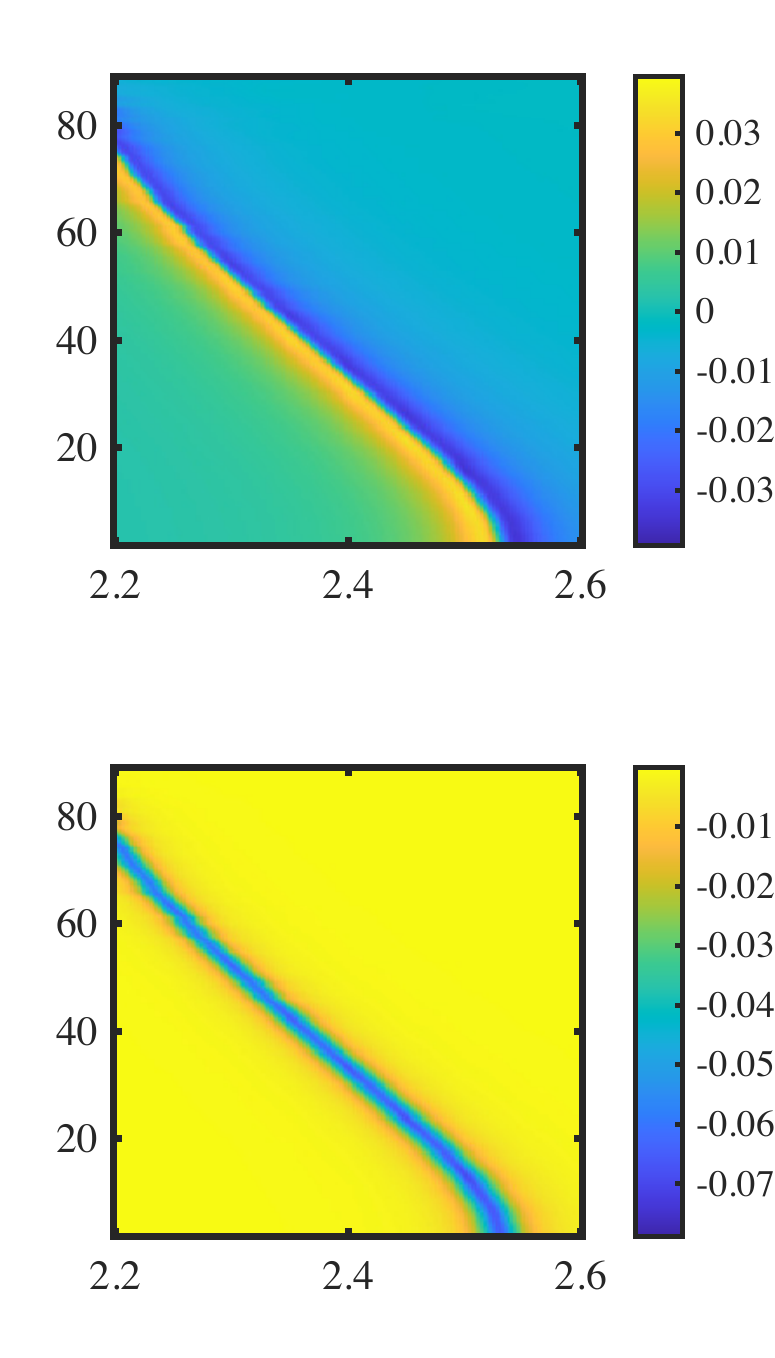}
				\put(1, 24){\vtext{\scriptsize  Incidence Angle, $\theta$~(deg)}}
				\put(1, 76){\vtext{\scriptsize  Incidence Angle, $\theta$~(deg)}}
				\put(25, 52){\htext{\scriptsize  Wire Length, $\ell$ (mm)}}
				\put(25, 1){\htext{\scriptsize  Wire Length, $\ell$ (mm)}}
				\put(25, 97){\htext{\color{amber} \scriptsize  Re$\{\chia{ee}{zz}\}$, FEM-HFSS}}
				\put(25, 46){\htext{\color{amber}\scriptsize  Im$\{\chia{ee}{zz}\}$, FEM-HFSS}}
	\end{overpic} \caption{}
\end{subfigure}	
\hfill
\begin{subfigure}{0.475\columnwidth}
				\begin{overpic}[width=\columnwidth,grid=false,trim={0cm 0cm 0cm 0cm},clip]{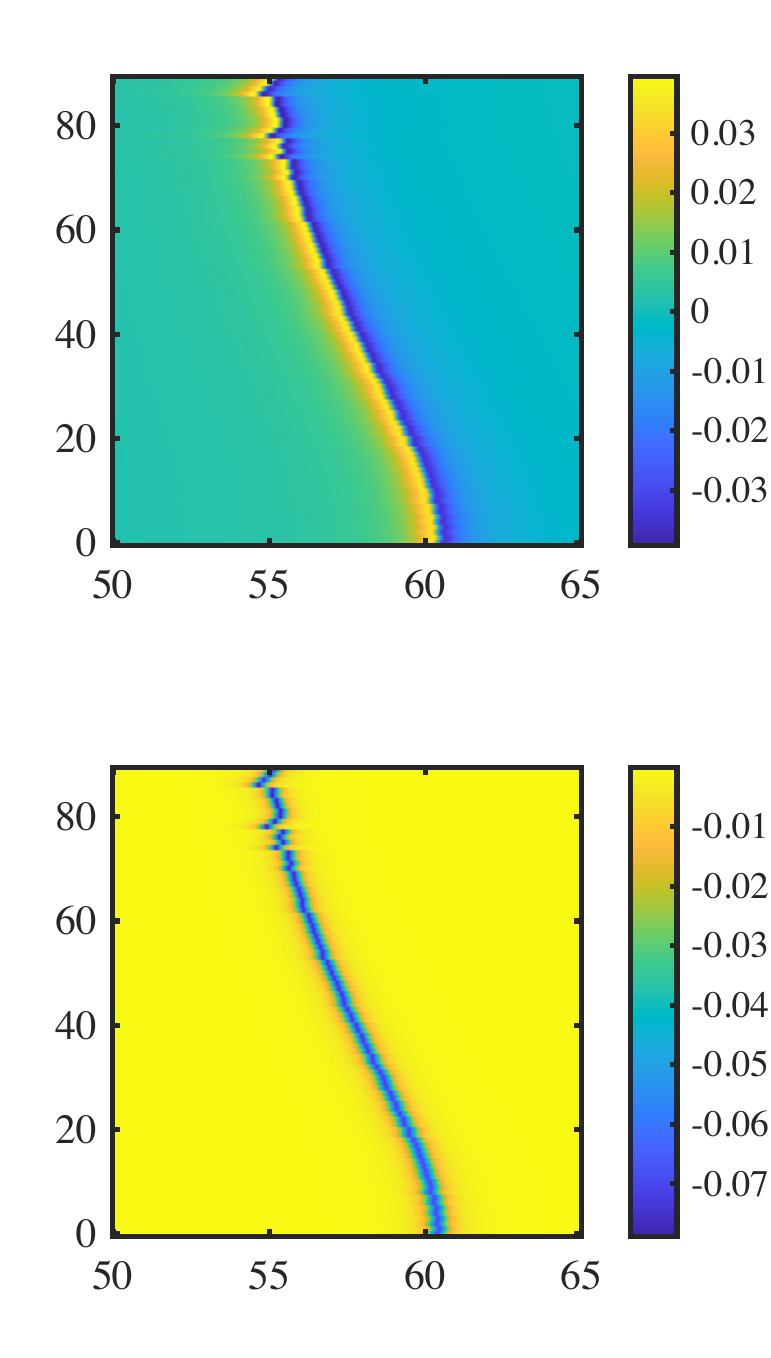}
				\put(1, 24){\vtext{\scriptsize  Incidence Angle, $\theta$~(deg)}}
				\put(1, 76){\vtext{\scriptsize  Incidence Angle, $\theta$~(deg)}}
				\put(25, 52){\htext{\scriptsize  Frequency (GHz)}}
				\put(25, 1){\htext{\scriptsize  Frequency (GHz)}}
				\put(25, 97){\htext{\color{amber} \scriptsize  Re$\{\chia{ee}{zz}\}$, FEM-HFSS}}
				\put(25, 46){\htext{\color{amber}\scriptsize  Im$\{\chia{ee}{zz}\}$, FEM-HFSS}}
	\end{overpic} \caption{}
\end{subfigure}	
\hfill
\begin{subfigure}{0.475\columnwidth}
				\begin{overpic}[width=\columnwidth,grid=false,trim={0cm 0cm 0cm 0cm},clip]{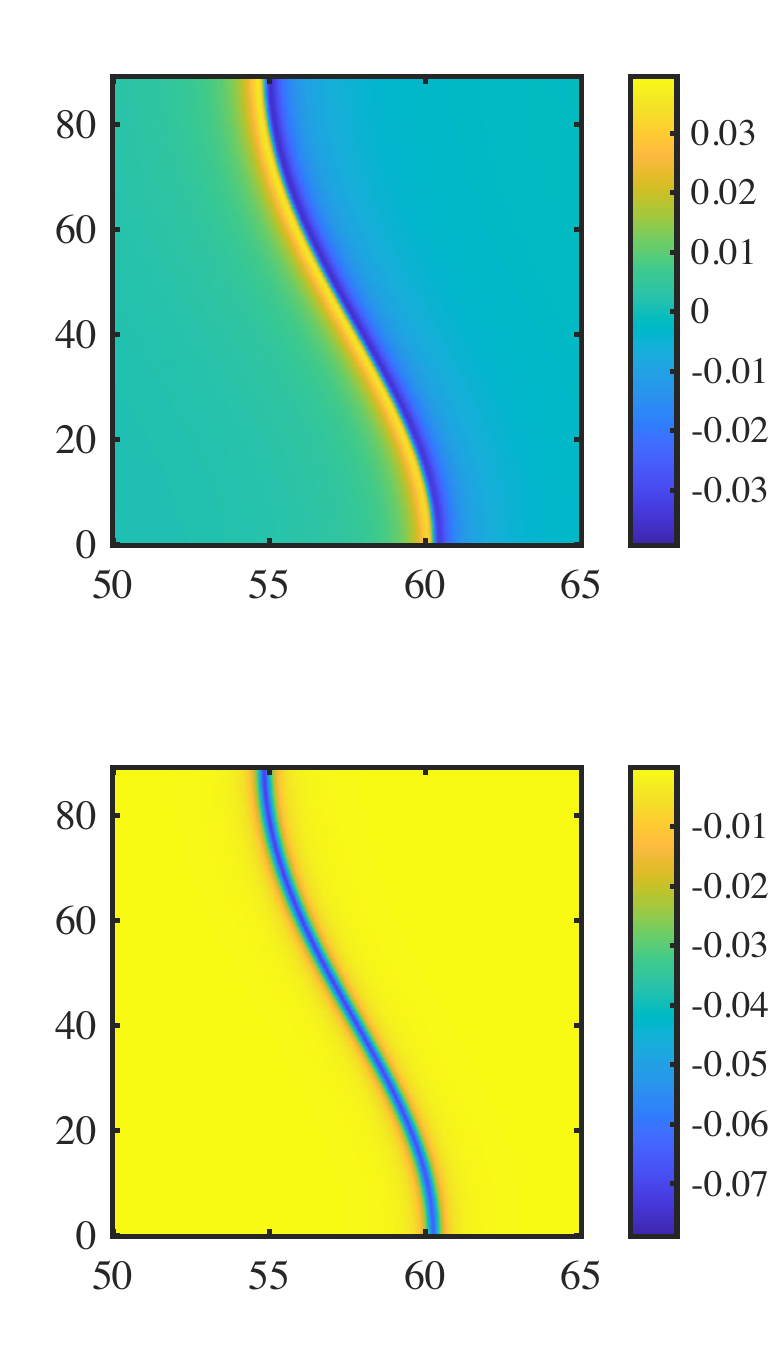}
				\put(1, 24){\vtext{\scriptsize  Incidence Angle, $\theta$~(deg)}}
				\put(1, 76){\vtext{\scriptsize  Incidence Angle, $\theta$~(deg)}}
				\put(25, 52){\htext{\scriptsize  Frequency (GHz)}}
				\put(25, 1){\htext{\scriptsize  Frequency (GHz)}}
				\put(25, 97){\htext{\color{ao} \scriptsize  Re$\{\chia{ee}{zz}\}$, Lorentz, Eq.~\eqref{Eq:Lor_kx_w_2}}}
				\put(25, 46){\htext{\color{ao}\scriptsize  Im$\{\chia{ee}{zz}\}$, Lorentz, Eq.~\eqref{Eq:Lor_kx_w_2}}}
	\end{overpic} \caption{}
\end{subfigure}	
\\
\vspace{0.2cm}
\begin{subfigure}{1.9\columnwidth}
				\begin{overpic}[width=\columnwidth,grid=false,trim={0cm 0cm 0cm 0cm},clip]{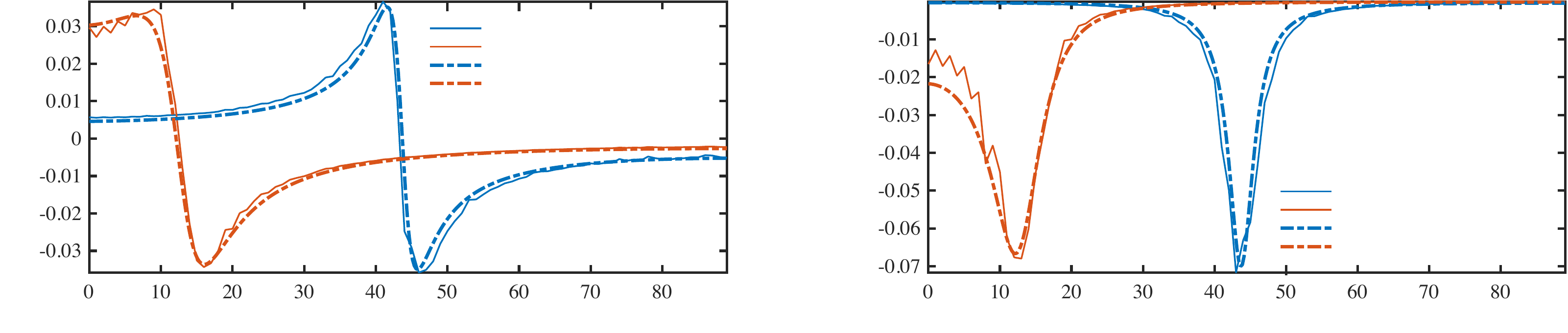}
	\put(25, 0){\htext{\scriptsize  Incidence Angle, $\theta$~(deg)}}
	\put(80, 0){\htext{\scriptsize  Incidence Angle, $\theta$~(deg)}}
	\put(1, 12){\vtext{\scriptsize  Re$\{\chia{ee}{zz}\}$}}
	\put(54, 12){\vtext{\scriptsize  Im$\{\chia{ee}{zz}\}$}}
	\put(34.5, 19.25){\htext{\tiny  HFSS (58 GHz)}}	
	\put(34.5, 18){\htext{\tiny  HFSS (60 GHz)}}	
	\put(35, 16.85){\htext{\tiny  Eq.~\eqref{Eq:Lor_kx_w_2} (58 GHz)}} 
	\put(35, 15.65){\htext{\tiny  Eq.~\eqref{Eq:Lor_kx_w_2} (60 GHz)}}
	\put(88.7, 8.85){\htext{\tiny  HFSS (58 GHz)}}	
	\put(88.7, 7.6){\htext{\tiny  HFSS (60 GHz)}}	
	\put(89.2, 6.45){\htext{\tiny  Eq.~\eqref{Eq:Lor_kx_w_2} (58 GHz)}} 
	\put(89.2, 5.25){\htext{\tiny  Eq.~\eqref{Eq:Lor_kx_w_2} (60 GHz)}}
	\end{overpic} \caption{}
\end{subfigure}	
				\caption{The angle-dependent surface susceptibilities for the short metal dipole. Simulation parameters are: $\sigma=580$ kS/m, $\ell=2.5$ mm, $d_0=0.2$ mm, $\Lambda_y = 2.15$ mm and $\Lambda_z= 4.3$ mm. (a) Unit cell configuration. (b) Electric susceptibilities at 60GHz for varying wire length, $\ell$. (c) Broadband electric susceptibilities as a function of angle from Ansys FEM-HFSS. (d) Surface fitting with parameters shown in Tab. \ref{Tab: DipoleParameters}. Due to symmetry, angular sweep in one space quadrant only is shown. (e) Comparison between  FEM-HFSS (solid) and Lorentz fitting (dot-dashed) at 58 and 60 GHz.}
		\label{fig:WireExample}
\end{figure*}

Let us consider the first example of a short electric dipole unit cell formed using a conducting wire, as shown in Fig.~\ref{fig:WireExample}(a), excited with a TE mode ($E_z,~H_y,~H_x$). It is simulated in Ansys FEM-HFSS using Floquet boundary conditions, where its transmittance and reflectance are computed for a sweeping angle of plane-wave incidence, $\theta$, using \eqref{Eq:TangChi}. At 60 GHz and for different wire lengths, $\ell$, Fig.~\ref{fig:WireExample}(b) shows that the spatial resonance is located at normal incidence ($\theta=0^\circ$) when $\ell=2.5$ mm, and it moves to higher angles as the wire length decreases.  For a fixed length $\ell=2.5$ mm,  Fig.~\ref{fig:WireExample}(c) shows a typical surface susceptibility distribution as a function of both temporal frequency, $\omega$ and spatial frequency $k_y$. The magnetic susceptibility is found to be negligible and thus not shown. It is clear that the surface susceptibilities are strongly angular dependent, and thus the resulting metasurface is expected to be spatially dispersive. Moreover, we observe a strong resonance migrating towards lower temporal frequencies for increasingly oblique angles.

\begin{table}[]
\fontsize{7pt}{10pt}\selectfont
\centering
\caption{Short Electric Dipole: Lorentz Resonator Properties}
\label{Tab: DipoleParameters}
\setlength{\tabcolsep}{1pt}
\begin{tabular}{CCCCCCC}
\toprule
\multicolumn{7}{c}{Tangential Electric Susceptibility $({\chia{ee}{zz}})$} \\
\bottomrule
   \multicolumn{1}{c}{$\alpha_0$} & \multicolumn{1}{c}{$\alpha_2$} & \multicolumn{1}{c}{$\beta_0$} & \multicolumn{1}{c}{$\beta_2$} & \multicolumn{1}{c}{$\zeta_0$} & \multicolumn{1}{c}{$\zeta_2$} & \multicolumn{1}{c}{$\chi_{\text{ee}_0}^{zz}$} \\
   \left(\frac{\text{G}\text{rad}}{\text{s}}\right) & \left(\frac{\text{rad}\cdot\text{m}^2}{\text{s}}\right) & \left(\frac{\text{G}\text{rad}}{\text{s}}\right) & \left(\frac{\text{P}\text{rad}^2\cdot\text{m}^2}{\text{s}^2}\right) &\left(\frac{\text{G}\text{rad}}{\text{s}}\right)& \left(\frac{\text{P}\text{rad}^2\cdot\text{m}^2}{\text{s}^2}\right)& \left(\times10^{-6}\right)
   \\ \midrule
 2.4677     & 419.6       & 7.712      & 12.5    & 378.69    & -18.59     & 0 \\ \bottomrule
\end{tabular}
\end{table}

Next, to capture the angle-dependent surface susceptibility, and in particular, a single angle-dependent resonance of the structure, the Lorentz oscillator model of \eqref{Eq:Lor_kx_w_2} is used to numerically curve-fit this response. Fig.~\ref{fig:WireExample}(c-d) shows the reconstructed electric and magnetic surface susceptibility profile across both temporal frequencies, $\omega$ and spatial frequencies $k_y = k_0\sin\theta$. There is a remarkable agreement between the full-wave simulated susceptibility and the reconstructed one, despite noisy data from HFSS, possibly due to poor convergence specially at higher angles. It thus confirms that the Lorentz model with only six non-zero parameters (see Tab. \ref{Tab: DipoleParameters}) fully describes such a complex response of this unit cell. Consequently, the extended GSTCs for this structure are simply given by \eqref{Eq:SpatialLorentz} with $\chi_\mm^{yy} = 0$.

\subsection{Huygens' Metasurface}

\begin{figure}[tbp]
		\centering
		\begin{subfigure}{\columnwidth}
					\begin{overpic}[width= 0.95\columnwidth,grid=false,trim={0cm 0cm 0cm 0cm},clip]{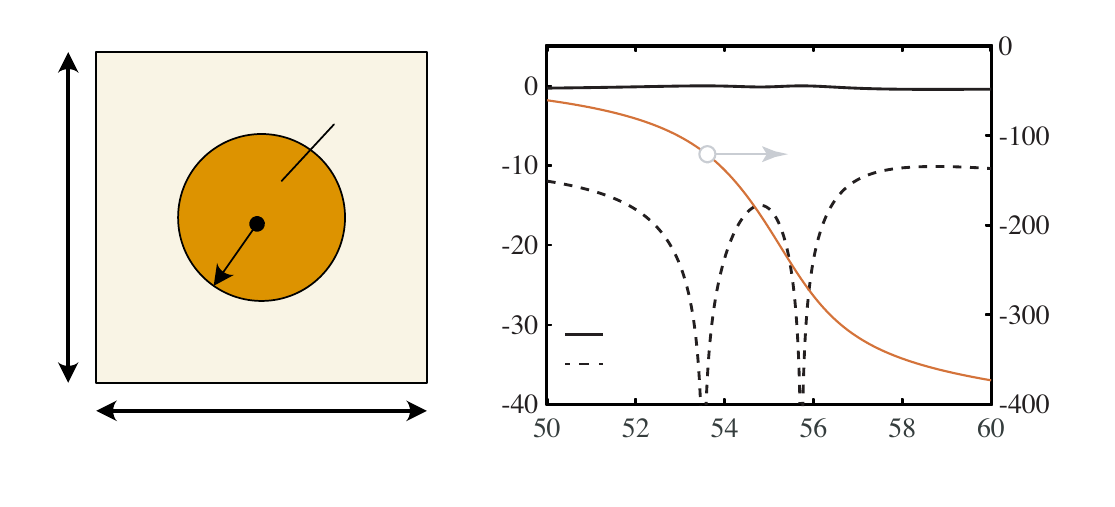}
					\put(50, 45){\htext{\footnotesize \color{amber}\shortstack{\textbf{\textsc{Dielectric Puck}} Unit Cell, $\chia{ee}{zz},~\chia{mm}{yy}$}}}
				\put(22, 4){\htext{\scriptsize  $\Lambda$}} 
				\put(3, 25){\htext{\scriptsize  $\Lambda$}}
				\put(25, 22){\htext{\scriptsize  $r_0$}}
				\put(29, 37){\htext{\scriptsize  Dielectric, $\epsilon_r$}}
				\put(25, 13){\htext{\scriptsize  Free-space, $\epsilon_h=1$}}
				
				\put(70, 2.5){\htext{\scriptsize  Frequency (GHz)}}
				\put(42, 24){\vtext{\scriptsize  S-parameters (dB)}}
				\put(96, 24){\rotatebox[origin=c]{-90}{\scriptsize  Phase $\angle S_{21}$~(deg)}}
				\put(58, 15){\htext{\tiny $|S_{21}|$}}
				\put(58, 12.5){\htext{\tiny $|S_{11}|$}}
		\end{overpic} \caption{}
\end{subfigure}
\\ \vspace{0.2cm}
\begin{subfigure}{\columnwidth}
				\begin{overpic}[width=0.95\columnwidth,grid=false,trim={0cm 0cm 0cm 0cm},clip]{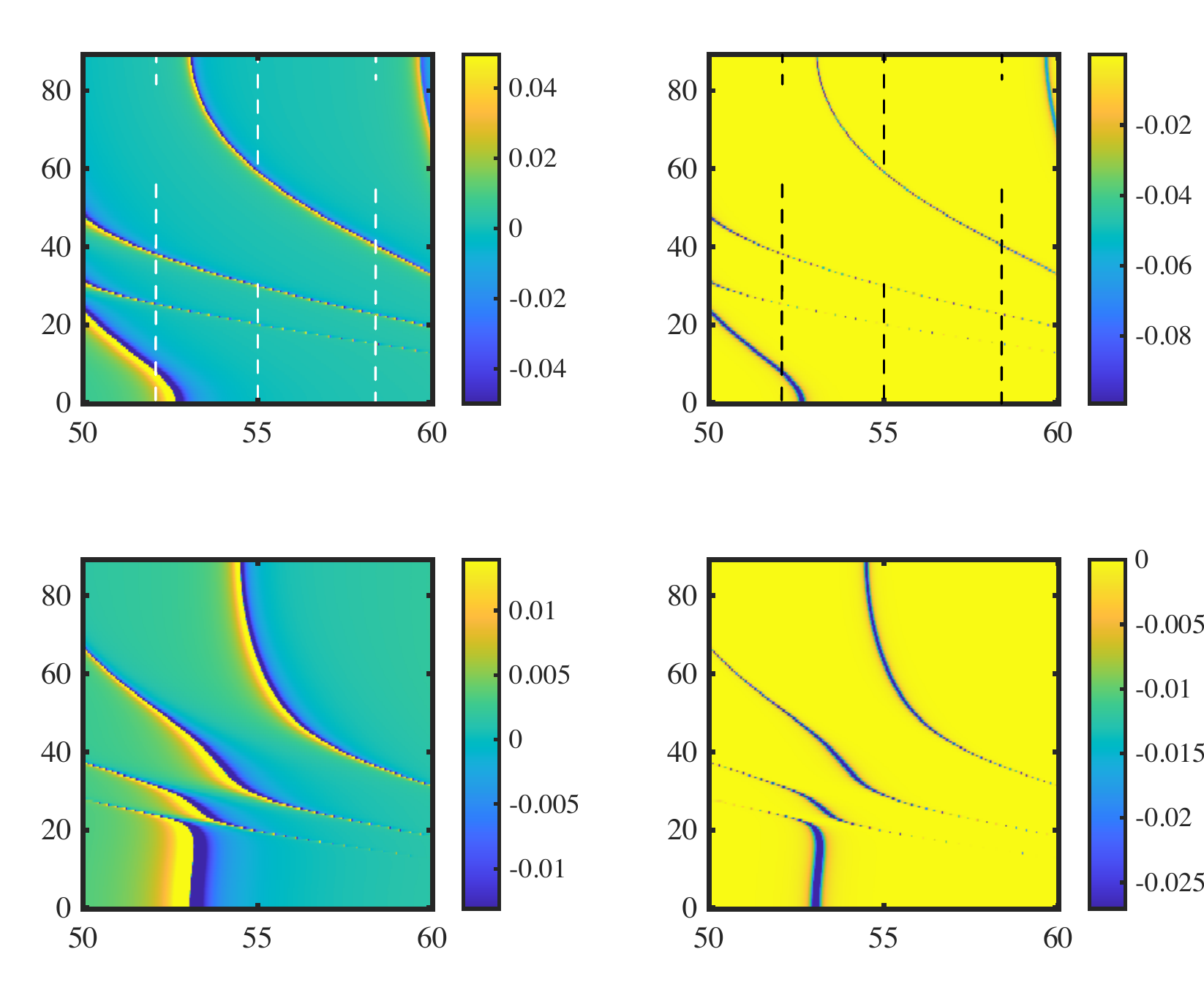}
				\put(75, 1.5){\htext{\scriptsize  Frequency (GHz)}}\put(75, 43.5){\htext{\scriptsize  Frequency (GHz)}}
				\put(22, 1.5){\htext{\scriptsize  Frequency (GHz)}}\put(22, 43.5){\htext{\scriptsize  Frequency (GHz)}}
				\put(0, 20){\vtext{\scriptsize  Incidence Angle, $\theta$~(deg)}}
				\put(0, 62){\vtext{\scriptsize  Incidence Angle, $\theta$~(deg)}}
				\put(22, 81){\htext{\color{amber}\scriptsize  Re$\{\chia{ee}{zz}$\}}}\put(75, 81){\htext{\color{amber}\scriptsize  Im$\{\chia{ee}{zz}$\}}}
				\put(22, 38.5){\htext{\color{amber}\scriptsize  Re$\{\chia{mm}{yy}$\}}}\put(75, 38.5){\htext{\color{amber}\scriptsize  Im$\{\chia{mm}{yy}$\}}}
				\put(12.5, 70.75){\vtext{\tiny  52~GHz}}
				\put(21, 63){\vtext{\tiny  55~GHz}}
				\put(30.75, 70.5){\vtext{\tiny  58~GHz}}
				\put(64.5, 70.75){\vtext{\tiny  52~GHz}}
				\put(73, 63){\vtext{\tiny  55~GHz}}
				\put(82.75, 70.5){\vtext{\tiny  58~GHz}}
	\end{overpic} \caption{}
\end{subfigure}	
\\ \vspace{0.2cm}
\centering
		\begin{subfigure}{\columnwidth}
					\begin{overpic}[width=0.95\columnwidth,grid=false,trim={0cm 0cm 0cm 0cm},clip]{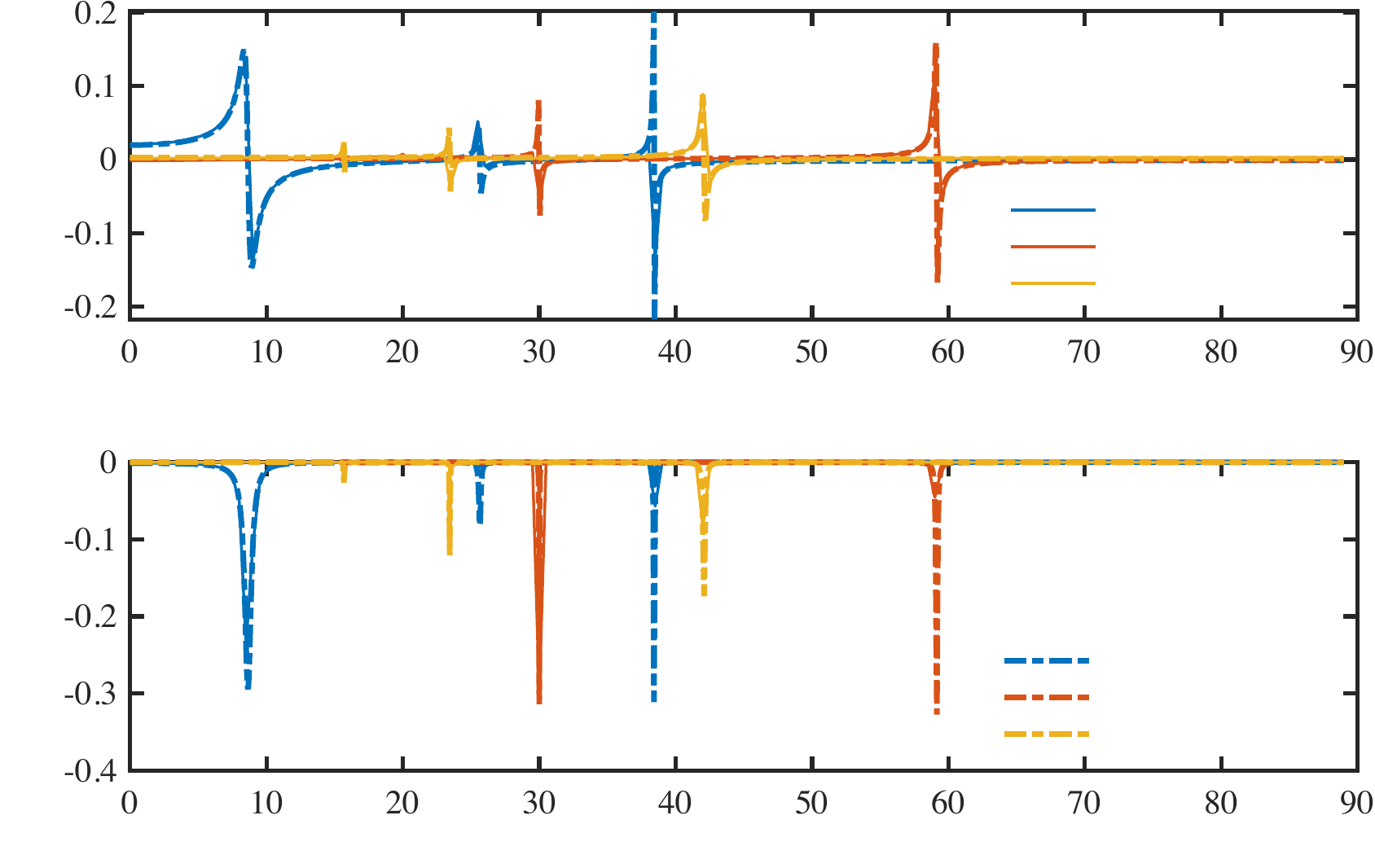}	
					\put(55, 0){\htext{\scriptsize  Incidence Angle, $\theta$~(deg)}}
					\put(1, 18){\vtext{\scriptsize  Im$\{\chia{ee}{zz}\}$}}\put(1, 50.5){\vtext{\scriptsize  Re$\{\chia{ee}{zz}\}$}}
						\put(87.75, 47.25){\htext{\tiny  HFSS (52 GHz)}}	\put(87.75, 44.5){\htext{\tiny  HFSS (55 GHz)}}	\put(87.75, 41.75){\htext{\tiny  HFSS (58 GHz)}}
	\put(88, 14.5){\htext{\tiny  Eq.~\eqref{Eq:LorentzSum} (52 GHz)}} 
	\put(88, 11.75){\htext{\tiny  Eq.~\eqref{Eq:LorentzSum} (55 GHz)}} 
	\put(88, 9){\htext{\tiny  Eq.~\eqref{Eq:LorentzSum} (58 GHz)}}
		\end{overpic}
		\\
		\\
    	\begin{overpic}[width=0.95\columnwidth,grid=false,trim={0cm 0cm 0cm 0cm},clip]{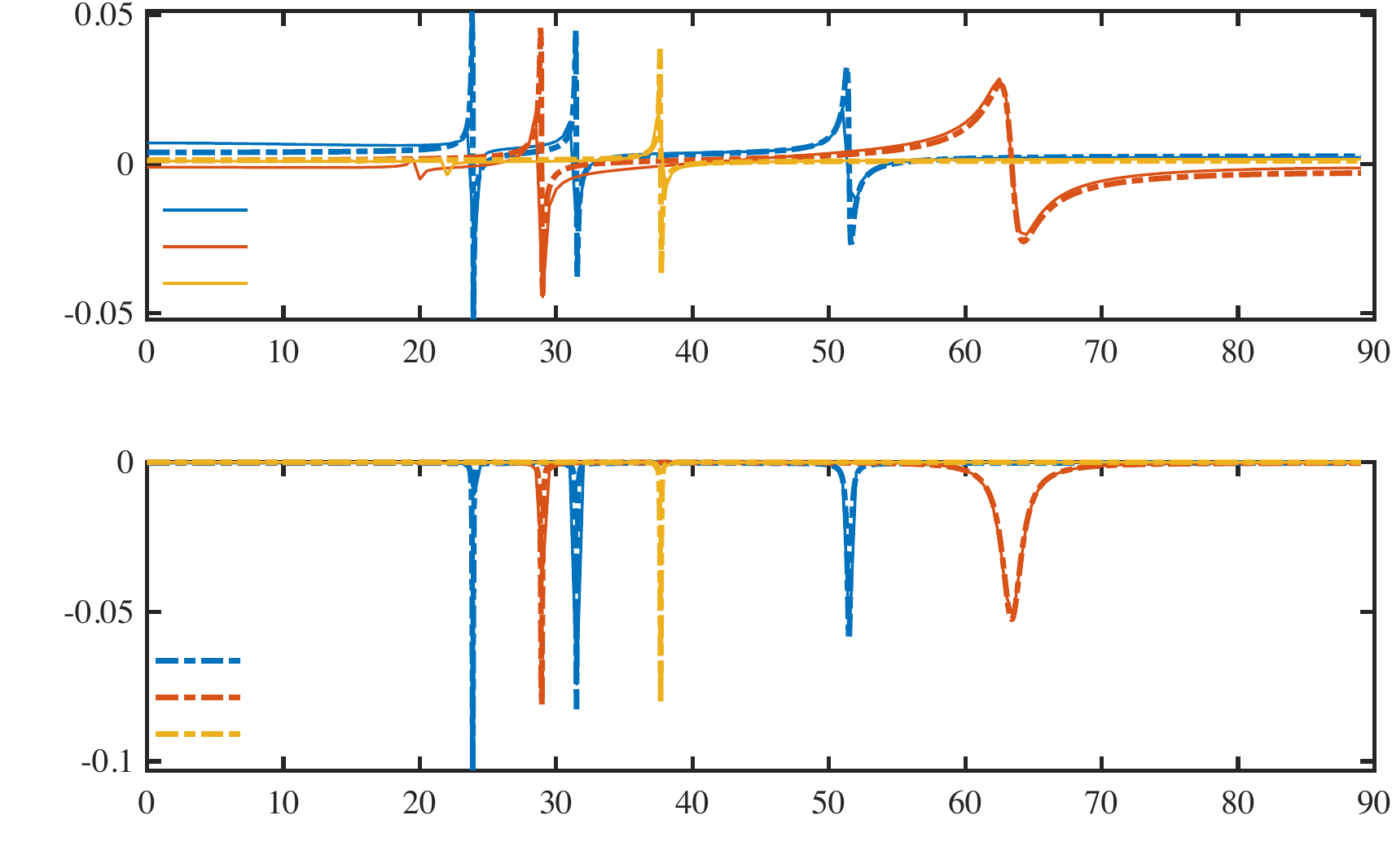}
	\put(55, 0){\htext{\scriptsize  Incidence Angle, $\theta$~(deg)}}
	\put(1, 18){\vtext{\scriptsize  Im$\{\chia{mm}{yy}\}$}}\put(1, 50.5){\vtext{\scriptsize  Re$\{\chia{mm}{yy}\}$}}
	\put(25.75, 47){\htext{\tiny  HFSS (52 GHz)}}	
	\put(25.75, 44.25){\htext{\tiny  HFSS (55 GHz)}}	
	\put(25.75, 41.5){\htext{\tiny  HFSS (58 GHz)}}
	\put(25.8, 14.5){\htext{\tiny  Eq.~\eqref{Eq:LorentzSum} (52 GHz)}}
	\put(25.8, 11.75){\htext{\tiny  Eq.~\eqref{Eq:LorentzSum} (55 GHz)}} 
	\put(25.8, 9){\htext{\tiny  Eq.~\eqref{Eq:LorentzSum} (58 GHz)}}
	\end{overpic} \caption{}
\end{subfigure}	
		\caption{Angle-dependent surface susceptibilities of a dielectric puck. a) Unit cell configuration. Simulation parameters are: $\epsilon_r = 6.5$, $r_0=1.4$~mm, $h=1.27$~mm, $\Lambda = 4.2$~mm. b) Broadband electric and magnetic surface susceptibilities as a function of angle, extracted from Ansys FEM-HFSS. c) Fitted susceptibilities with Lorentzian functions at 52, 55 and 58 GHz. Fitting parameters are presented in Tab. \ref{Tab: PuckParameters}.}
		\label{fig:DieelctricPuck}
\end{figure}

Next, consider an all-dielectric resonator structure, which consists of a cylindrical dielectric puck made of high permittivity material embedded inside a host medium of lower permittivity (assumed air for simplicity here), as shown in Fig.~\ref{fig:DieelctricPuck}(a). All-dielectric structures are common, especially at optical frequencies as Huygens' structures (co-located orthogonal electric and magnetic dipoles), due to their low-loss characteristics and their zero backscattering property as further shown in Fig.~\ref{fig:DieelctricPuck}(a) \cite{AllDieelctricMTMS}. They have also been proposed at millimeter-wave frequencies in both all-dielectric \cite{Emara_Huygens, Emara_Huygens_CP, EmaraCP} and standard printed circuit board implementations \cite{Soichi_Huygen_mmWave}. One common feature among all these structures is their relatively large unit cell sizes which can approach free-space wavelength, making them weakly sub-wavelength.

The typical angle-dependent electric and magnetic surface susceptibilities of a dielectric puck are shown in Fig.~\ref{fig:DieelctricPuck}(b), simulated in FEM-HFSS, and susceptibilities extracted using \eqref{Eq:TangChi}. Compared to the simpler unit cells of Fig.~\ref{fig:WireExample}, the dielectric unit cell features a more complicated angular dependence of the susceptibilities. Specifically, it shows multiple angular resonances across the angular spectrum drifting across the temporal frequencies, which suggests that a single Lorentzian oscillator model of \eqref{Eq:Lor_kx_w_2} is insufficient to model this structure. While one can use a brute force approach to fit this response using rational polynomials of the form of \eqref{Eq:Ratio}, we observe that at each temporal frequency, surface susceptibilities appear as a summation of several angular resonances with Lorentz characteristics. Consequently, we next develop a multi-Lorentz surface description of this structure.

Let us assume that the electric surface polarization components (and analogous development for the magnetic ones) are described using $N_L$ Lorentz resonators and that, without loss of generality, each resonator's properties are up to second-order dependent on $k_y$. Then, \eqref{Eq:Lor_kx_w_2} assumes the form,
\begin{align}\label{Eq:LorentzSum}
\Ptf_z &=  \Ptf_{z_0} + \sum_{i=1}^{N_L} \Ptf_{z,i} \notag\\
&= \epsilon_0\chi_{\ee_0}^\zz\Etf_{z,\text{av}} + \sum_{i=1}^{N_L}  \epsilon_0 \frac{\chi_{\ee,0,i}^\zz+j\chi_{\ee,1,i}^\zz k_y-\chi_{\ee,2,i}^\zz k_y^2}{(1 + j\xi_{\ee,1,i}^\zz k_y - \xi_{\ee,2,i}^\zz k_y^2)}\Etf_{z,\text{av}},
\end{align}
where the constant term, $\chi_{\ee_0}^\zz$ is added for generality. Taking the inverse spatial Fourier transform, $\mathcal{F}_y^{-1}\{\cdot\}$ on each side, we get
\begin{align*}
\sum_{i=1}^{N_L} \left[ \xi_{\ee,2,i}^\zz \frac{\partial^2}{\partial y^2} +  \xi_{\ee,1,i}^\zz\frac{\partial}{\partial y}
+ 1\right]\Pt_{z,i} = \epsilon_0\chi_{\ee_0}^\zz\Et_{z,\text{av}}\notag\\
 + \epsilon_0\sum_{i=1}^{N_L} \left[ \chi_{\ee,0,i}^\zz +\chi_{\ee,1,i}^\zz \frac{\partial}{\partial y}+\chi_{\ee,2,i}^\zz \frac{\partial^2}{\partial y^2}\right]\Et_{z,\text{av}}.
\end{align*}
Furthermore, from the GSTCs $\Delta\Ht_{y,i} = j\omega \Pt_{z,i}$ and using superposition and the independence of the polarizations, we retrieve the extended form of the GSTCs for this case as
\begin{align*}
    \Delta\Ht_y &= \Delta \Ht_{y_0} + \sum_{i=1}^{N_L} \Delta \Ht_{y,i} = j\omega\Pt_{z_0} + \sum_{i=1}^{N_L} j\omega \Pt_{z,i}
\end{align*}
\noindent where the $0^\text{th}$ order term is,
\begin{align*} 
\Delta \Ht_{y_0} = j\omega\epsilon_0\chi_{\ee_0}^\zz\Et_{z,\text{av}}
\end{align*}
\noindent and for each term $0<i \leq N_L$,
\begin{align*}
& \left[\xi_{\ee,2,i}^\zz \frac{d^2}{dy^2} +  \xi_{\ee,1,i}^\zz\frac{d}{dy} + 1\right]\Delta \Ht_{y,i} = \notag\\
 & j\omega\epsilon_0\left[ \chi_{\ee,0,i}^\zz +\chi_{\ee,1,i}^\zz \frac{\partial}{\partial y}+\chi_{\ee,2,i}^\zz \frac{\partial^2}{\partial y^2}\right]\Et_{z,\text{av}}.
\end{align*}
The analogous development for the magnetic surface polarization leads to
\begin{align*}
    \Delta \Et_{z_0} = j\omega\chi_{\mm_0}^\yy\Ht_{y,\text{av}}
\end{align*}
\noindent and,
\begin{align*}
& \left[\xi_{\mm,2,i}^\yy \frac{d^2}{dy^2} +  \xi_{\mm,1,i}^\yy\frac{d}{dy} + 1\right]\Delta \Et_{z,i} = \notag\\
 & j\omega\left[ \chi_{\mm,0,i}^\yy +\chi_{\mm,1,i}^\yy \frac{\partial}{\partial y}+\chi_{\mm,2,i}^\yy \frac{\partial^2}{\partial y^2}\right]\Ht_{y,\text{av}}.
\end{align*}
It should be noted that a sum of Lorentzians \eqref{Eq:LorentzSum} may equivalently be seen as a ratio of two polynomials of \eqref{Eq:Ratio} when all combined, which in general may lead to spatial derivatives higher than 2. However, their decomposition in terms of Lorentzians not only provides a physically motivated description, it allows us to write the spatial boundary conditions involving spatial derivatives up to second-order only, which is more suitable for numerical implementation.

To demonstrate the multi-Lorentz description of the all-dielectric cell, Fig.~\ref{fig:DieelctricPuck}(c) shows the electric and magnetic surface susceptibilities at couple of arbitrarily chosen frequencies and as a function of angle. An excellent fitting is observed across the entire angular spectrum, confirming the suitability of \eqref{Eq:LorentzSum} to model this unit cell. Moreover, a complete description of the surface polarizabilities requires several Lorentz resonators to capture the whole angular response at discrete frequencies. Data in Tab~\ref{Tab: PuckParameters} presents the values of the parameters for each frequency. It further reinforces the advantage of the presented methodology in describing spatial dispersion using a compact set of parameters.

\begin{table}[]
\fontsize{7pt}{10pt}\selectfont
\centering
\caption{Dielectric Puck: Lorentz Resonators Properties}
\label{Tab: PuckParameters}
\setlength{\tabcolsep}{1pt}
\begin{tabular}{CCCCCCCCC}
\toprule
\multicolumn{9}{c}{Tangential Electric Susceptibility $({\chia{ee}{zz}})$} \\
\bottomrule
   &   & \multicolumn{1}{c}{$\alpha_0$} & \multicolumn{1}{c}{$\alpha_2$} & \multicolumn{1}{c}{$\beta_0$} & \multicolumn{1}{c}{$\beta_2$} & \multicolumn{1}{c}{$\zeta_0$} & \multicolumn{1}{c}{$\zeta_2$} & \multicolumn{1}{c}{$\chi_{\text{ee}_0}^{zz}$} \\
   & & \multicolumn{1}{c}{$\left(\frac{\text{M}\text{rad}}{\text{s}}\right)$} & \multicolumn{1}{c}{$\left(\frac{\text{rad}\cdot\text{m}^2}{\text{s}}\right)$} & \multicolumn{1}{c}{$\left(\frac{\text{G}\text{rad}}{\text{s}}\right)$} & \multicolumn{1}{c}{$\left(\frac{\text{P}\text{rad}^2\cdot\text{m}^2}{\text{s}^2}\right)$} &\multicolumn{1}{c}{$\left(\frac{\text{G}\text{rad}}{\text{s}}\right)$}& \multicolumn{1}{c}{$\left(\frac{\text{P}\text{rad}^2\cdot\text{m}^2}{\text{s}^2}\right)$}& \multicolumn{1}{c}{$\left(\times10^{-3}\right)$}
   \\ \midrule
\multirow{3}{*}{\rotatebox{90}{\tiny{52 GHz}}} 	& 1 	& 	827.73 	& 	879.99 	& 	9.07 	& 	0.00 	& 	333.30 	& 	-162.06 	& 	\multirow{3}{*}{$-0.08 -j0.02$} \\
 	& 2 	& 	86.43 	& 	1847.21 	& 	3.80 	& 	0.00 	& 	370.52 	& 	-137.36 	& 	 \\
 	& 3 	& 	-16.65 	& 	472.14 	& 	5.38 	& 	0.00 	& 	404.80 	& 	-124.56 	& 	 \\ \midrule
\multirow{2}{*}{\rotatebox{90}{\tiny{55 GHz}}} 	& 1 	& 	-5.09 	& 	176.44 	& 	2.41 	& 	0.00 	& 	377.37 	& 	-69.24 	& 	\multirow{2}{*}{$0.76$} \\
 	& 2 	& 	37.83 	& 	415.24 	& 	7.19 	& 	0.00 	& 	499.58 	& 	-132.99 	& 	 \\ \midrule
\multirow{3}{*}{\rotatebox{90}{\tiny{55 GHz}}} 	& 1 	& 	25.06 	& 	676.08 	& 	1.27 	& 	0.00 	& 	383.68 	& 	-133.41 	& 	\multirow{3}{*}{$1.91 -j0.02$} \\
 	& 2 	& 	-37.81 	& 	724.56 	& 	2.47 	& 	0.00 	& 	404.01 	& 	-130.19 	& 	 \\
 	& 3 	& 	350.40 	& 	535.16 	& 	6.71 	& 	0.00 	& 	465.94 	& 	-127.08 	& 	 \\ \toprule
\multicolumn{9}{c}{Tangential Magnetic Susceptibility $({\chia{mm}{yy}})$} \\ \bottomrule
   &   & \multicolumn{1}{c}{$\alpha_0$} & \multicolumn{1}{c}{$\alpha_2$} & \multicolumn{1}{c}{$\beta_0$} & \multicolumn{1}{c}{$\beta_2$} & \multicolumn{1}{c}{$\zeta_0$} & \multicolumn{1}{c}{$\zeta_2$} & \multicolumn{1}{c}{$\chi_{\text{mm}_0}^{yy}$} \\
   & & \multicolumn{1}{c}{$\left(\frac{\text{M}\text{rad}}{\text{s}}\right)$} & \multicolumn{1}{c}{$\left(\frac{\text{rad}\cdot\text{m}^2}{\text{s}}\right)$} & \multicolumn{1}{c}{$\left(\frac{\text{G}\text{rad}}{\text{s}}\right)$} & \multicolumn{1}{c}{$\left(\frac{\text{P}\text{rad}^2\cdot\text{m}^2}{\text{s}^2}\right)$} &\multicolumn{1}{c}{$\left(\frac{\text{G}\text{rad}}{\text{s}}\right)$}& \multicolumn{1}{c}{$\left(\frac{\text{P}\text{rad}^2\cdot\text{m}^2}{\text{s}^2}\right)$}& \multicolumn{1}{c}{$\left(\times10^{-3}\right)$}
   \\ \midrule
   
   \multirow{3}{*}{\rotatebox{90}{\tiny{52 GHz}}} 	& 1 	& 	215.77 	& 	0.00 	& 	2.84 	& 	0.00 	& 	367.40 	& 	-145.00 	& 	\multirow{3}{*}{$3.12 -j0.26$} \\
 	& 2 	& 	327.46 	& 	0.00 	& 	2.99 	& 	0.00 	& 	384.09 	& 	-125.91 	& 	 \\
 	& 3 	& 	718.36 	& 	0.00 	& 	3.75 	& 	0.00 	& 	403.66 	& 	-77.34 	& 	 \\ \midrule
\multirow{2}{*}{\rotatebox{90}{\tiny{55 GHz}}}	& 1 	& 	402.19 	& 	0.00 	& 	3.73 	& 	-2.97 	& 	398.19 	& 	-125.89 	& 	\multirow{2}{*}{$0.79 -j0.04$} \\
 	& 2 	& 	5681.95 	& 	0.01 	& 	3.49 	& 	85.61 	& 	501.82 	& 	-124.70 	& 	 \\ \midrule
\multirow{2}{*}{\rotatebox{90}{\tiny{58 GHz}}} 	& \multirow{2}{*}{$1$} 	& 	\multirow{2}{*}{$360.92$} 	& 	\multirow{2}{*}{$0.10$} 	& 	\multirow{2}{*}{$3.24$} 	& 	\multirow{2}{*}{$0.18$} 	& 	\multirow{2}{*}{$467.35$} 	& 	\multirow{2}{*}{$-155.29$} 	& 	\multirow{2}{*}{$1.08 -j0.01$} \\ 
 \\ \bottomrule

\end{tabular}
\end{table}

\section{Spatial Dispersion vs Normal Surface Polarizations}

\begin{figure}[tbp]
		\centering
		\begin{subfigure}[b]{\linewidth}
		\centering
       \begin{overpic}[width=0.7\linewidth,grid=false,trim={0cm 0cm 0cm 0cm},clip]{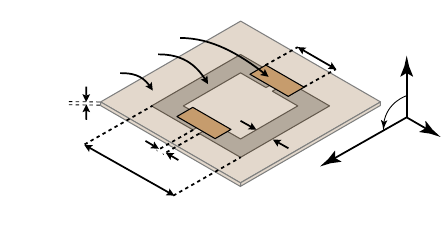}
				\put(10, 31){\htext{\scriptsize 0.025}}
				\put(23, 14){\htext{\scriptsize 2.5}}
				\put(22, 42){\htext{\scriptsize bottom copper}}
				\put(30, 47){\htext{\scriptsize top copper}}
				\put(20, 38){\htext{\scriptsize PI film}}
				\put(32, 18){\htext{\scriptsize 0.2}}
				\put(55, 30){\htext{\scriptsize 0.4}}
				\put(75, 45){\htext{\scriptsize $\ell$}}
				\put(5,5){\scriptsize \textit{(Units are mm)}}
				\put(73.5,27){\htext{\scriptsize $\Lambda$}}
				\put(68, 16){\htext{\scriptsize $y$}}
				\put(101, 22){\htext{\scriptsize $z$}}
				\put(91, 44){\htext{\scriptsize $x$}}
				\put(87, 22){\htext{\scriptsize $\theta$}}
				\put(20, 55){\htext{\footnotesize \color{amber}\shortstack{\textbf{\textsc{Loop Resonator Unit Cell}}\\ $\chib{ee}{zz}$,~$\chib{mm}{xx}$}}}
			\end{overpic}\caption{}
	 \end{subfigure}
\\\vspace{0.2cm}
\begin{subfigure}{\columnwidth}
				\begin{overpic}[width=\columnwidth,grid=false,trim={0cm 0cm 0cm 0cm},clip]{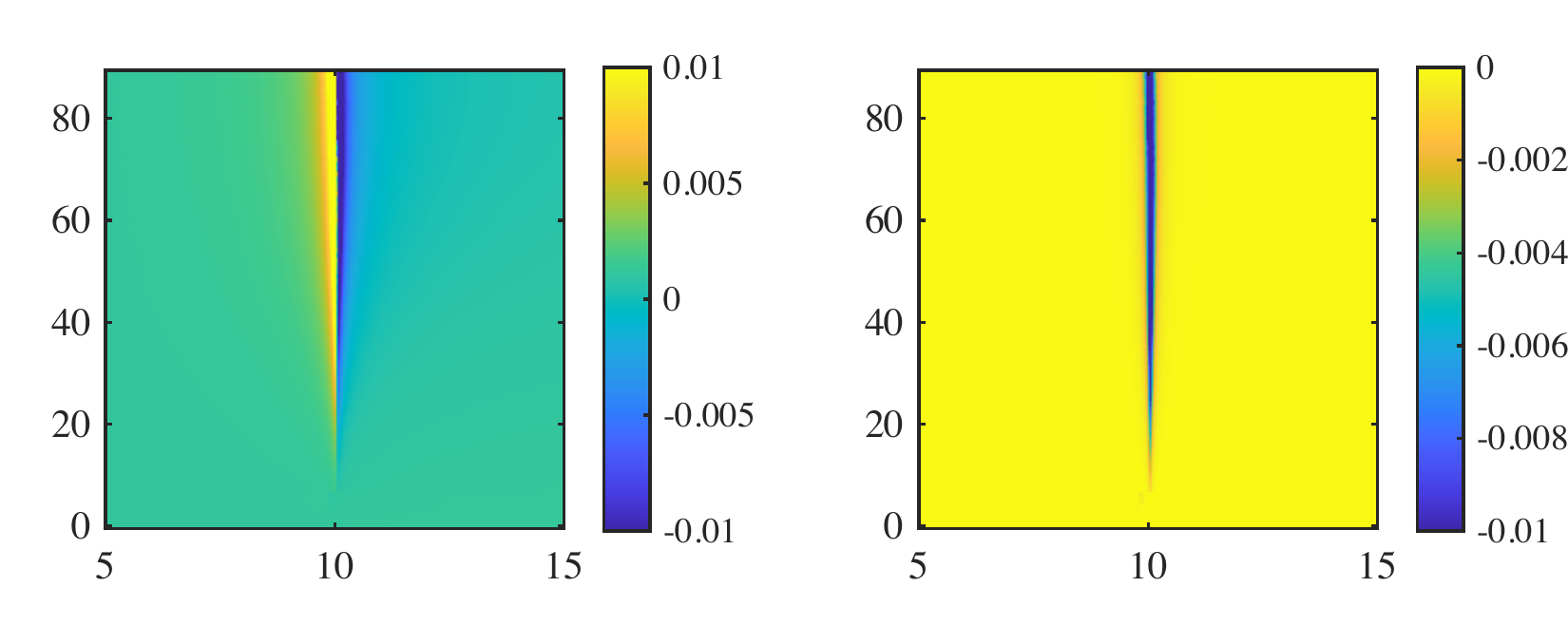}
				\put(0, 21){\vtext{\scriptsize  Incidence Angle, $\theta$~(deg)}}\put(53, 21){\vtext{\scriptsize  Incidence Angle, $\theta$~(deg)}}
				\put(23, 1){\htext{\scriptsize  Frequency (GHz)}}\put(75, 1){\htext{\scriptsize  Frequency (GHz)}}
				\put(21, 39){\htext{\color{amber} \scriptsize  Re$\{\chia{ee}{zz}\}$, FEM-HFSS}}
				\put(75, 39){\htext{\color{amber}\scriptsize  Im$\{\chia{ee}{zz}\}$, FEM-HFSS}}
	\end{overpic} \caption{}
\end{subfigure}	
\\\vspace{0.2cm}
\begin{subfigure}{0.95\columnwidth}

			\begin{overpic}[width=\columnwidth,grid=false,trim={0cm 0cm 0cm 0cm},clip]{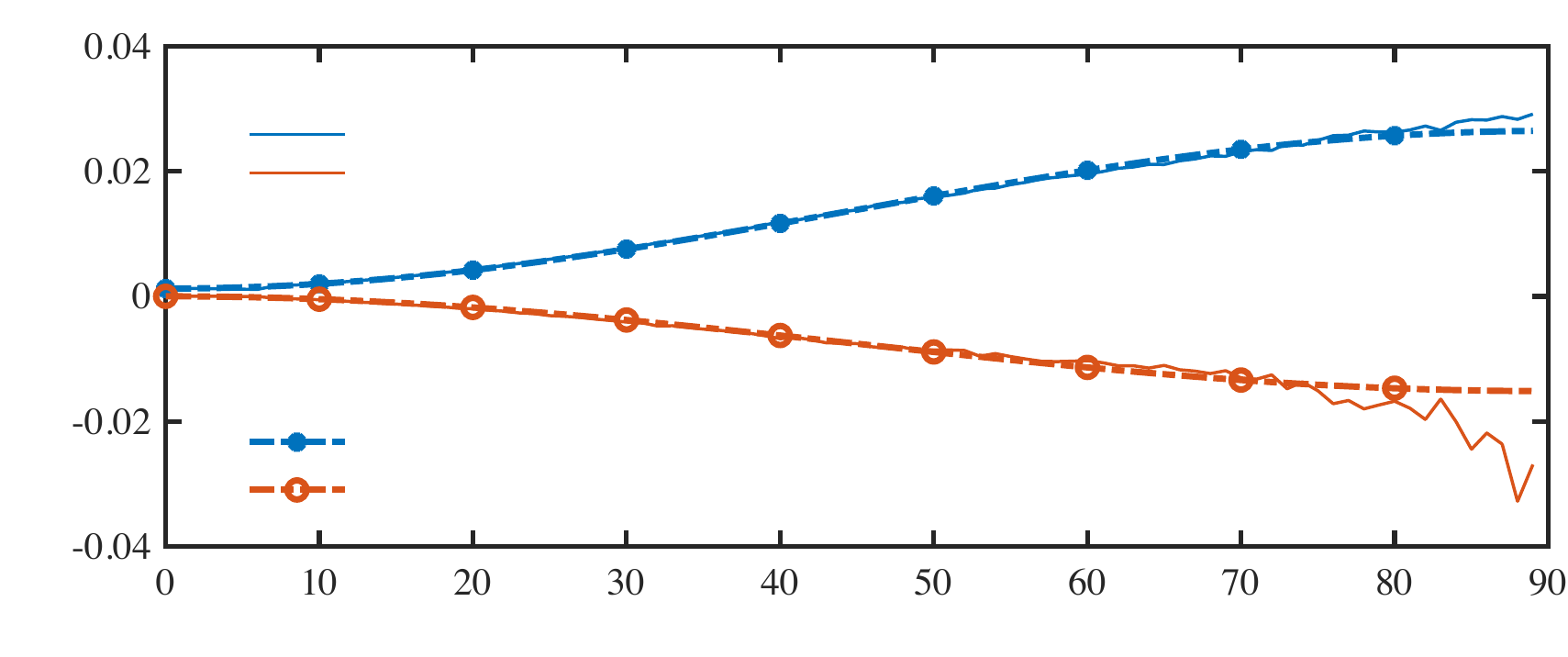}
			\put(50, 0){\htext{\scriptsize  Incidence Angle, $\theta$~(deg)}}
			\put(2, 22){\vtext{\scriptsize  $\{\chia{ee}{zz}\}$}}
			\put(31, 13){\htext{\tiny  Re$\{\cdot\}$, Eq.~\eqref{Eq:NormwithSD}}}
			\put(31, 10){\htext{\tiny  Im$\{\cdot\}$, Eq.~\eqref{Eq:NormwithSD}}}
			\put(30, 33){\htext{\tiny  Re$\{\cdot\}$, Eq.~\eqref{Eq:TangChi}}}
			\put(30, 30){\htext{\tiny  Im$\{\cdot\}$, Eq.~\eqref{Eq:TangChi}}}
	\end{overpic} \caption{}
\end{subfigure}	
		\caption{Angle dependent surface susceptibilities of a sub-wavelength unit cell. a) Unit cell configuration. b) Broadband equivalent tangential electric surface susceptibility as a function of angle. c) Equivalence between susceptibilities calculated using Eq. \eqref{Eq:TangChi} and \eqref{Eq:NormwithSD} at 10 GHz.}
		\label{fig:NormalExample}
\end{figure}
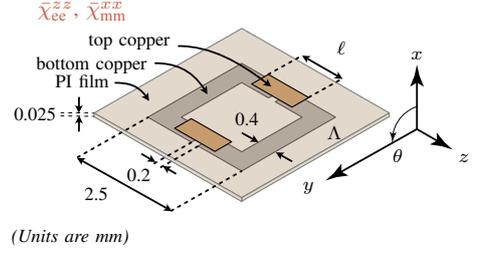
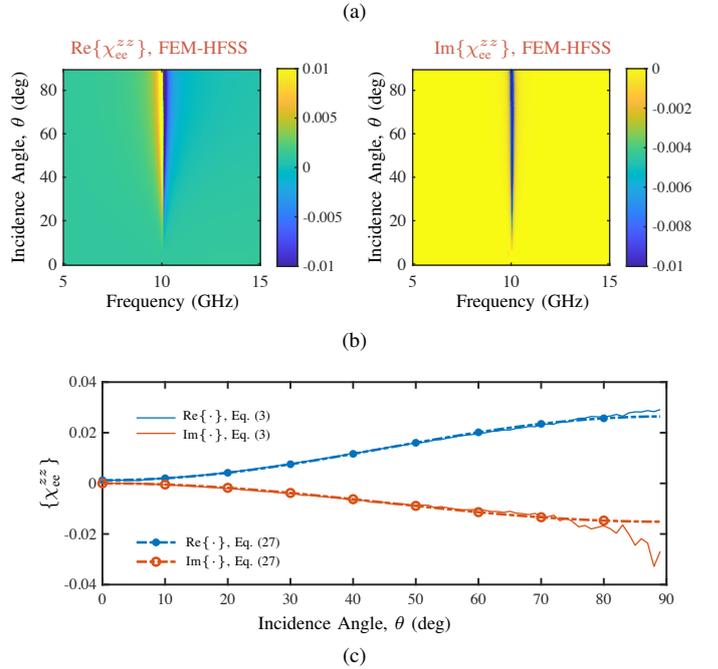

So far, we have considered spatially dispersive structures which exhibit angle-dependent \emph{tangential} surface susceptibilities only and showed how they lead to extended GSTCs, which can describe the angular scattering from the metasurfaces. In a variety of other structures which are spatially non-dispersive, the angular scattering is described using both the normal and the tangential surface susceptibility components, dictated by their respective physical mechanisms. For example, consider a resonant loop structure consisting of a Metal-Insulator-Metal (MIM) capacitor printed on a thin dielectric slab, as shown in Fig.~\ref{fig:NormalExample}(a). At oblique incidence, the time-varying magnetic flux through the loop induces an electric current around the conducting loop via Faraday's law, which leads to a strong normal magnetic polarization along the $x-$axis. Consequently, it has been shown that the angular scattering of such a unit cell structure can be accurately modeled using one tangential surface susceptibility, $\chia{ee}{zz}$ and one normal susceptibility component $\chia{mm}{xx}$, with \emph{no spatial dispersion}~\cite{smy2021iegstc, VilleFloq}. 

As seen, spatial dispersion and normal surface susceptibility components are related to determining the angular scattering from the surface. Therefore, one may wonder if they are related to each other or represent two independent properties of a given metasurface? For instance, is it possible to model the unit cell structure of Fig.~\ref{fig:NormalExample}(a) using purely tangential surface susceptibilities which are angle-dependent and thus spatially dispersive? To answer this question, we recall that the transmittance and reflectance of a metasurface described using one tangential surface susceptibility, $\chib{ee}{zz}$ and one normal susceptibility component $\chib{mm}{xx}$, are given by
\begin{subequations}\label{Eq:RT_Normal}
\begin{align}
R_\perp &= -\frac{jk_0\{\sin^2\theta \chib{mm}{xx}+\chib{ee}{zz}\}}{\{jk_0\sin^2\theta \chib{mm}{xx} + jk_0\chib{ee}{zz}+2\cos\theta\}}\\
T_\perp &= \frac{2\cos\theta}{\{jk_0\sin^2\theta \chib{mm}{xx} + jk_0\chib{ee}{zz}+2\cos\theta\} }
\end{align}
\end{subequations}
where $\theta$ is the angle of incidence of an incoming plane-wave (TE mode), and $k_0$ is the free-space wavenumber. On the other hand, the transmittance and reflectance of a metasurface described using a single tangential surface susceptibility, $\chia{ee}{zz}$, and \emph{no normal component}, are given by
\begin{subequations}\label{Eq:RT_Tang}
\begin{align}
R_{||} &= -\frac{jk_0 \chia{ee}{zz}}{\{jk_0\chia{ee}{zz}+2\cos\theta\} }\\
T_{||} &= \frac{2\cos\theta}{\{jk_0\chia{ee}{zz}+2\cos\theta\} }
\end{align}
\end{subequations}
Comparing \eqref{Eq:RT_Normal} and \eqref{Eq:RT_Tang}, it is clear that for $R_\perp = R_{||}$ and $T_\perp = T_{||}$ for every angle of incidence $\theta$, we must have, 
\begin{align}\label{Eq:NormwithSD}
\chia{ee}{zz} &= \chib{ee}{zz} + \left(\frac{\chib{mm}{xx}}{k_0^2}\right)k_y^2,
\end{align}
where the effect of normal component $\chib{mm}{xx}$ has been absorbed in the new tangential component $\chia{ee}{zz}$. To confirm this equivalence, Fig.~\ref{fig:NormalExample}(b) shows the extracted $\chia{ee}{zz}$ directly from FEM-HFSS using \eqref{Eq:TangChi}, which shows an angle-independent resonance frequency, but whose plasma frequency varies with the angle of incidence (as seen from the line-width broadening of the resonances). Furthermore, Fig.~\ref{fig:NormalExample}(c) presents the comparison between the extracted susceptibilities at 10~GHz and the analytical relation of a fictitious $\chia{ee}{zz}$ obtained from \eqref{Eq:NormwithSD}. A very good fit is observed, confirming this equivalence for a uniform plane-wave incidence.

While this demonstration may suggest that a spatially non-dispersive surface with normal surface susceptibilities may be represented using a spatially dispersive metasurface with tangential surface susceptibilities only, this equivalence is not generally correct. It is, in fact, applicable for \emph{a uniform metasurface only} and can be explained by considering the GSTCs for the case of a general non-uniform metasurface. Specifically, for a metasurface described using tangential $\chib{ee}{zz}$ and normal $\chib{mm}{xx}$, we get,
\begin{align}\label{Eq:GSTCs_tang_norm}
\Delta H_y  = j\omega \epsilon_0\chib{ee}{zz}E_{z,\text{av}} -  \chib{mm}{xx}\frac{\partial H_{x,\text{av}}}{\partial y}   -  H_{x,\text{av}}\frac{\partial \chib{mm}{xx}}{\partial y}  
\end{align}
On the other hand, the GSTC for a metasurface with a single tangential $\chia{ee}{zz}$ described using \eqref{Eq:NormwithSD}, reads:
\begin{align}\label{Eq:GSTCtangonly}
\Delta H_y &= j\omega \epsilon_0 \chib{ee}{zz}E_{z,\text{av}}  -   \left\{\frac{\epsilon_0 \chib{mm}{xx}}{k_0^2} \right\} \frac{\partial^2E_{z,\text{av}} }{\partial y^2}
\end{align}
It is clear that \eqref{Eq:GSTCs_tang_norm} is only equal to \eqref{Eq:GSTCtangonly}, if $\partial \chib{mm}{xx}/\partial y= 0$, i.e. a uniform metasurface. Therefore, we can conclude that for a general nonuniform metasurface, spatial dispersion \emph{and} the normal surface susceptibilities (if they physically exist) represent two \emph{different} properties which must be taken into account simultaneously to accurately describe its complete angular scattering.

\section{Conclusions}

A simple method to describe spatially dispersive metasurfaces has been proposed where angle-dependent surface susceptibilities are explicitly used to formulate the zero thickness sheet model of practical metasurface structures. It has been shown that if the surface susceptibilities can be expressed as a ratio of two polynomials of tangential spatial frequencies  $k_{||}$, that captures their zero and pole behaviors, they can be conveniently expressed as spatial derivatives of the difference and average fields around the metasurface in the space domain. They thus represent extended GSTCs accounting for spatial dispersion. Using two simple examples of a short electric dipole and an all-dielectric cylindrical puck unit cells, which exhibit purely tangential surface susceptibilities and symmetric transmission and reflection responses, the proposed concept is numerically confirmed in 2D. A single Lorentzian has been found to describe the spatio-temporal frequency behavior for the short dipole, while a multi-Lorentzian description was necessary to capture the multiple angular resonances of the dielectric puck. In both cases, the appropriate spatial boundary conditions have been provided. We further emphasize that the proposed approach models very complex spatio-temporal responses of the two unit cells considered here, using very few parameters, thereby making them ideal compact simulation models of such structures to be easily integrated in standard electromagnetic field solvers.

As has been shown, the generalized expressions of the surface susceptibilities as a ratio of two polynomials in $k_y$ or as a sum of Lorentzian oscillators reveal themselves as spatial derivatives of the fields around the metasurface, resulting in more general boundary conditions than standard GSTCs, that can be incorporated in a variety of numerical methods to compute the scattered fields from spatially dispersive metasurfaces. For example, Part 2 of this work will demonstrate this integration of extended GSTCs in an Integral Equation (IE) based field solver, where the scattered fields may be computed in the temporal frequency domain for an arbitrary incidence wave and a given metasurface configuration. 

Moreover, in this work, the analysis has been limited to structures exhibiting purely tangential surface susceptibilities. However, it has been shown that to capture the complete angular scattering properties of a general metasurface, both spatial dispersion, and normal surface susceptibilities must be taken into account if the structure physically supports them. While the proposed method can be extended to include the normal components, developing the method to integrate them will be an important step. Last but not the least, general modeling of spatially dispersive non-uniform metasurfaces will be a natural extension where either the complex coefficients of \eqref{Eq:Ratio} or the parameters of the Lorentz oscillator become function of space. This work marks the first essential step to achieve this goal.

\section*{Acknowledgements}

The authors acknowledge funding from the Department of National Defence's Innovation for Defence Excellence and Security (IDEaS) Program in support of this work.

\balance
\bibliographystyle{ieeetran}
\bibliography{PREP_2021_MS_Spatial_Dispersion_TAP}  

\end{document}